\documentclass[a4paper]{article}

\usepackage{amsmath}
\DeclareSymbolFont{AMSa}{U}{msa}{m}{n}
\DeclareSymbolFont{AMSb}{U}{msb}{m}{n}
\DeclareSymbolFontAlphabet{\mathbb}{AMSb}

\newcommand{\rr}{\mathbb{R}}            

\newcommand{\kk}{\boldsymbol{\kappa}}            

\newcommand{\bfe}{\mathbf{e}}        
\newcommand{\bfE}{\mathbf{E}}        

\newcommand{\bfn}{\mathbf{n}}        

\newcommand{\bfg}{\mathbf{g}}        

\newcommand{\bfk}{\mathbf{k}}        

\newcommand{\Div}{{\sf{div}\,}}
\newcommand{\Det}{{\sf{det}\,}}
\newcommand{\Curl}{{\sf{curl}\,}}
\newcommand{\Grad}{{\sf{grad}\,}}

\newcommand{\bfchi}{\boldsymbol{
  \chi}}                             

\newcommand{\ljump}{[\mkern-2.6mu[}        
\newcommand{\rjump}{]\mkern-2.6mu]}        

\begin{document}

\title{ Multi-Phase Equilibrium\\ of  Crystalline Solids  }

\author{
        Paolo Cermelli   \\
        Dipartimento di Matematica \\
        Universit\`a di Torino  \\
        I--10123 Torino, Italy   \\
         { cermelli{@}dm.unito.it}
\and
        Shaun Sellers \\
        School of Mathematics \\
      University of East Anglia,\\
        Norwich NR4 7TJ, England,  \\
         { sellers@uea.ac.uk} }

\date{\small Key words: A.~phase transformation, A.~microstructures, A.~dislocations, \\
      B.~crystal plasticity, C.~variational calculus \\[.35in] 
}

\maketitle

\begin{abstract}   
A continuum model of crystalline solid equilibrium
 is presented in which the
underlying periodic lattice structure is taken explicitly into account. This
model also allows for
both point and line defects in the bulk of the lattice
and at interfaces, and the kinematics of such defects is discussed 
in some detail. 
A Gibbsian variational argument  is used to derive the necessary bulk and
interfacial conditions for  multi-phase equilibrium  (crystal-crystal and
crystal-melt) where the allowed lattice variations
involve the  creation and transport of
defects in the bulk and at the phase interface.
An interfacial energy, assumed to depend on the 
interfacial dislocation density and the orientation of the interface with 
respect to the lattices of both phases, is also included in the analysis.
Previous equilibrium results based on nonlinear elastic models 
for incoherent and coherent interfaces
are recovered as special cases
for when the lattice distortion is constrained to
coincide  with the macroscopic deformation gradient, thereby excluding
bulk dislocations. 
The formulation is purely spatial and needs no recourse to a fixed
reference configuration or
an elastic-plastic decomposition of the strain.  
Such a decomposition can be
introduced however through an incremental elastic deformation superposed onto
an already dislocated state, but leads to additional equilibrium conditions. 
The presentation emphasizes the role of {configurational
forces} as they provide a natural framework for
the description and interpretation of singularities and phase
transitions.
\end{abstract}


\section{Introduction}

Consider two phases of a given material separated by a sharp interface
and in equilibrium in a heat bath. Let one phase be a crystalline solid,
the other phase may be either the corresponding fluid
melt or another crystalline solid (such as in a twinned crystal).
This paper is concerned with the
 determination of the equilibrium conditions for the bulk regions
occupied by the two phases and for the interface separating them---a 
problem of significant interest in diverse fields. Numerous authors have
already studied various aspects of it and have derived
equations governing the mechanical, thermal, and chemical equilibrium
(Eshelby, 1970; Robin, 1974; Larch\'e \& Cahn, 1978,1985;
Cahn, 1980, Cahn \& Larch\'e, 1982;
Grinfel'd, 1981; Mullins, 1984; Alexander \& Johnson,
1985; Johnson \& Alexander, 1986; Kaganova \& Roitburd, 1988;
Leo \& Sekerka, 1989; Gurtin, 1993;
Cermelli \& Gurtin, 1994; Leo \& Hu, 1995).
A fundamental characteristic of a crystalline solid is its
lattice structure, in which the atoms are periodically spaced.
Additionally the lattice contains defects, e.g., point defects
(interstitials and vacancies)
and line defects (dislocations), both types of which can impact on the
equilibrium behavior. Larch\'e \& Cahn (1978) included point defects in their
original analysis by treating the interstitials and vacancies as additional
mobile species subject to a lattice or network constraint.
This work was later extended to allow
for surface energy associated with the interface
(Cahn, 1980, Cahn \& Larch\'e, 1982; Mullins, 1984;
Alexander \& Johnson, 1985; Johnson \&
Alexander, 1986; Leo \& Sekerka, 1989; Gurtin, 1993; Cermelli \&
Gurtin, 1994; Leo \& Hu, 1995). In this approach however
the lattice does not appear explicitly, so it is not clear
how other properties of the lattice, such as the distortion of the lattice or
the occurrence of
dislocations, affect the governing equations.

In this paper we present a continuum model of crystalline solid  equilibrium
that explicitly incorporates the underlying lattice structure
and includes the possibility of both point and line defects. 
Following
Larch\'e \& Cahn (1978), we then give for this model
a Gibbsian variational analysis, in which we obtain the necessary conditions
for two-phase crystal-crystal equilibrium.
We consider general non-equilibrium variations that allow for creation and
transport of defects
in the bulk and at the interface.  We treat  both coherent
and incoherent interfaces,
two extreme cases that characterize many types of interfaces
between crystalline solids. In particular, incoherency is important
since it can  arise from the presence of defects at the phase interface.
Additionally, we consider the case of
crystal-melt equilibrium.
We also include interfacial energy, assuming that it depends on the 
interfacial dislocation density and the orientation of the interface with 
respect to the lattices of both phases.

We model a crystalline solid by a continuum  microstructural
theory where the microstructural
variables represent in an average sense the underlying lattice structure:
\begin{itemize}
\item[-]3 linearly independent vector fields denote the \emph{average
lattice vectors}
at a spatial point;
\item[-]\emph{interstitials} and \emph{vacancies} are treated in the
usual way by modelling them as additional mobile
species subject to a lattice constraint;
\item[-]\emph{dislocations} are related to the anholonomicity (i.e.,
non-integrability) of the microstructural lattice vectors.
\end{itemize}
In this model we  introduce neither a fixed reference
configuration nor a macroscopic deformation gradient;
 rather we describe the state only by the microstructural
variables through a spatial description,
where quantities are defined on a per unit cell basis at a
fixed point in space.
A spatial description is uncommon in traditional models of solids, but
is more convenient when interfaces are incoherent as they
remain together in the current configuration.
It  also seems to be the most appropriate when relating to standard
x-ray observations of lattice points, which do not indicate how the
lattice points actually evolve.
We also assume that  the energy per unit cell of the current
configuration depends upon the local current  state  (i.e.,
lattice vectors, dislocation density, and occupancy  density).

Commonly in microstructural models, 
lattice vectors are assumed to deform as material
line elements---the so-called \emph{Born rule} (Ericksen, 1984)---so that the
lattice vectors in the deformed configuration are related to the
lattice vectors in the reference configuration by the macroscopic
deformation gradient. This assumption however is not always valid:
in particular it fails when there are dislocations;
it can also fail in other cases (Zanzotto, 1992). We therefore formulate
our model without employing the Born rule, but then we show how more
classical models for solids
are recovered when the Born rule is imposed as a constraint.
Thus our equilibrium results are new
in that
\begin{itemize}
\item[ (1)]
bulk and surface dislocations are explicitly taken into account; and

\item[  (2)] no \emph{a priori} relation between reference lattice vectors
and actual lattice vectors is assumed.

\end{itemize}
When the Born rule is imposed as a
constraint, the lattice vectors then deform as material line elements,
the  dislocation density vanishes in the bulk, the lattice distortion
becomes equivalent to the deformation gradient,
and we recover  equilibrium results equivalent to those of
Larch\'e \& Cahn (1978),
Leo \& Sekerka (1989), and Cermelli \& Gurtin (1994).

Our  approach, however, does not require any
recourse to the concept of macroscopic deformation or fixed reference
configuration.
On the other hand, for a number of applications or in certain special
cases, it may be convenient to introduce 
 the often used elastic-plastic decomposition
of the strain (Bilby \emph{et al}., 1957; Kr\"oner, 1960; Lee, 1969).
 We do so by identifying the dislocated crystal with the intermediate
configuration of such a decomposition 
and then superposing a classical elastic (i.e., defect-preserving) 
deformation  onto this
dislocated state. In this approach our ``elastic'' part of the elastic-plastic
decomposition is truly elastic in the traditional sense.
This procedure allows one
to construct a simple model of dislocated bodies while keeping track
of all the different elastic and plastic components
of the theory. Due to the additional structure in the theory arising from the
incremental elastic deformation, additional equilibrium conditions arise,
so that the resulting equilibrium theory is
\emph{not equivalent} to that without the elastic-plastic decomposition. 
In particular,
when such an intermediate configuration is introduced,
both the dislocated intermediate configuration and the current configuration
must be in mechanical equilibrium.

There is some precedence for a microstructural approach to
modelling crystalline solids.
Motivated by the Cosserat director theories for oriented materials
(Ericksen \& Truesdell, 1958; Truesdell \& Toupin, 1960),
both Fox (1966,1970) and Toupin (1968) proposed using a triad of continuous
vector fields to represent lattice vectors associated with a material point
and demonstrated how dislocations could be
related to the non-integrability of the directors.
The deformation associated with the lattice directors was assumed distinct
from the average, macroscopic deformation. Additionally Fox
 discussed how a notion of lattice slip could be introduced.
Mullins \& Sekerka (1985) also used  a triad
of vector fields to model the lattice vectors, which,
however, where taken to deform as material line
elements (the Born rule). Consequently  any
constitutive dependence on them could be replaced by a dependence on the
macroscopic deformation gradient,
which excluded the possibility of bulk dislocations.
In this case their model described anisotropic elastic solids (Ericksen \&
Rivlin, 1954).
Davini (1986) and  Davini \& Parry (1989,1991)
 discussed in detail the various defect measures
and their relation to the lattice directors. They also introduced
the neutral deformations---inelastic deformations that do not change the defect
content---and demonstrated a corresponding decomposition of the strain.
Naghdi \& Srinivasa (1993a,b,1994a,b) formulated a director
theory as  a model of  slip in plastic deformations.
Similarly, Besseling and van der Giessen (1994) introduced a triad of vector
fields for the lattice vectors
in their discussion of inelastic deformations, but did not
introduce the notion of defects in this model.
More recently,  D{\l}u\.zewski  (1996) discussed the driving forces acting
on defects in a director model.
And Ericksen (1997) discussed  an equilibrium  theory of crystals
in terms of lattice directors without
employing the Born rule and without introducing a macroscopic
deformation gradient, but did not consider defects. Our approach follows
closely that of Davini, Parry, and Ericksen in that it is based entirely
on a description in terms of the underlying lattice structure.

Finally,  our presentation emphasizes the role of the \emph{configurational
forces},\footnote{Also called \emph{Eshelby forces} or
\emph{energy-momentum}.}
as they provide a natural framework for studying singularities
and phase transitions
(Eshelby, 1951,1970; Maugin, 1993; Gurtin, 1995). Configurational forces
do work over changes in the reference lattice and
arise naturally in our theory as the derivatives of the energy with respect
to the reciprocal lattice vectors (Cermelli \& Sellers, 1998). In our
calculations the configurational forces result as primary quantities and
 provide an immediate
interpretation of the resulting equilibrium conditions.


\section{Crystalline Solid Model }

\subsection{Motivation}

By the expression crystalline solid, we mean a model for a solid that
 explicitly incorporates the underlying lattice structure typical of a crystal.
Our model of a defective crystalline solid is based on the notion that
the regular lattice structure is, on the average, locally
recognizable and measurable (Davini, 1986). For instance,
x-ray observations can
provide local values for the various lattice parameters.
Further, such experimental measurements indicate that
the dilatation of the lattice parameters does not always
correspond to the  average macroscopic
strain as measured in, for example, tensile tests. 
These results intimate that the  macroscopic deformation gradient
can be in some cases an \emph{inappropriate} measure of the true
deformation of the crystal;
finer kinematic details of the lattice may be needed. In the following we
introduce microstructural variables  to represent
such kinematic details of the lattice and its distortion. In this point
of view, the state of
a  crystalline solid with defects is
completely determined by the  crystalline
structure, which we take as  the set of lattice vectors and the occupancy
density of lattice sites. These quantities are viewed as  fields over
the region of space
occupied by the material in its actual or current configuration.

\subsection{Lattice microstructure}

Let $\mathcal{B}\subset \mathbb{R}^3$ denote the region of space occupied
by the crystalline solid
in its \emph{actual} or \emph{current}
configuration. We will consistently use this
spatial region
as a reference.

To each spatial point ${\bf x}\in\mathcal{B}$, we associate  a triad of vector
fields ${\bf e}_i({\bf x})$ (where $i$ ranges from 1 to 3) that represent
the \emph{lattice
basis vectors} at ${\bf x}$. They describe in an average sense
the local microscopic arrangement of the lattice points and vary
continuously from
one spatial point to  another. It is also useful to introduce the
\emph{reciprocal lattice vectors}
${\bf e}^i({\bf x})$ defined by  the relation
\begin{equation}
{\bf e}^i\!\cdot {\bf e}_j= \delta^i_j,
\qquad {\bf e}_i\otimes{\bf e}^i=
 {\bf e}^i \otimes{\bf e}_i={\bf 1}, \label{reciprocal_lattice}
\end{equation}
where  $\delta^i_j$ is the Kronecker delta symbol and repeated Latin
indices are
summed.
Generally x-ray observations provide direct measurements
of the reciprocal lattice vectors ${\bf e}^i$.
We shall also denote   a fixed, constant and uniform
lattice basis by ${\bf E}_{i}$, with reciprocal basis ${\bf E}^{i}$.
It provides a convenient reference lattice from which to measure
the change of the lattice vectors.\footnote{Although we do not introduce
a fixed reference configuration for the current configuration
$\mathcal{B}$, we do use the notion
of \emph{reference lattice}.}

The \emph{lattice distortion}
is by definition the tensor field
\begin{equation}
{\bf F}_{\rm lat}=
{\bf e}_{i}\otimes{\bf E}^{i},
\label{strains}
\end{equation}
which takes the reference lattice vectors ${\bf E}_{i}$ to the actual
lattice vectors ${\bf e}_{i}$ at each spatial point. Similarly, the
\emph{inverse lattice distortion}
\begin{equation}{\bf F}_{\rm lat}^{-1}=
{\bf E}_i\otimes{\bf e}^i
\label{inverse_strains}
\end{equation}
takes the actual lattice vectors to the reference lattice vectors
at each spatial point. We do not assume that
${\bf F}_{\rm lat}$ or $ {\bf F}_{\rm lat}^{-1}$ are gradients.
In particular, ${\bf F}_{\rm lat}$ need not correspond to the 
macroscopic deformation gradient ${\bf F}$.

Since  the reference lattice vectors are constant, the \emph{actual} or 
\emph{current state} of the lattice
is  equivalently characterized by either the fields
${\bf e}_i$, here taken as spatial fields on $\mathcal{B}$, or by the fields
${\bf e}^i$, due to the relation
(\ref{reciprocal_lattice}) between
lattice vectors and reciprocal lattice vectors.


%

Furthermore, we interpret ${\bf e}_1\cdot({\bf e}_2\times{\bf e}_3)$ as
the
volume of a unit cell, so that
\begin{equation}
 n:= {\bf e}^1\cdot({\bf e}^2\times{\bf e}^3)=
\frac{1}{ {\bf e}_1\cdot({\bf e}_2\times{\bf e}_3)}
\label{number_density}
\end{equation}
is the \emph{number of cells per unit volume}.
It is also a continuously varying spatial field defined on
 $\mathcal{B}$.

\subsection{Point defects}

In a perfect crystal the atoms occupy lattice points. In a defective
crystal some of the lattice points may not be occupied, these being called
\emph{vacancies}. Additionally, some atoms may be at points not located
at the lattice points, these being called \emph{interstitials}.
Here, to maintain the presentation simple, we assume that there exists a
single mobile species of atoms, either vacancies or interstitials.
The extension to an arbitrary number of
species is straightforward.
We thus introduce the following scalar fields:
\begin{alignat*}{2}
 &\varrho  ({\bf x}) && \qquad \text{\it atoms per unit cell}              \\
 &\varrho_{\rm vac} ({\bf x})  && \qquad  \text{\it vacancies per unit cell}
\end{alignat*}
If $\varrho_{\rm vac}>0$, we will interpret $\varrho_{\rm vac}$
as a density of vacancies; if $\varrho_{\rm vac}<0$,
we will interpret $|\varrho_{\rm vac}|$
as a density of interstitial atoms.
Since we allow for only a single mobile species,
only the sign of  $\varrho_{\rm vac}$ distinguishes between
vacancies and interstitials.
These  two fields are not independent but are related by
the \emph{lattice constraint}:
\begin{equation}
\varrho +\varrho_{\rm vac} =\ell = \mbox{const.},
\label{lattice_constraint}
\end{equation}
where $\ell$ is the number of lattice points per unit cell.
The lattice points are regarded as a geometric constant of the
lattice.
Consequently only one of the
quantities is independent, and we arbitrarily choose
$\varrho$ as the independent density.

If independent mobile point defects are allowed,
they must also be specified
in addition to the lattice vectors.
Thus for our model, the \emph{current state of a crystalline solid with
a single mobile species} is given by $({\bf e}_i, \varrho)$,
specified as spatial fields on $\mathcal{B}$.
Alternatively and equivalently,
 the lattice vectors ${\bf e}_i$ can be replaced
by the reciprocal lattice vectors ${\bf e}^i$ or the molar density $\varrho$
replaced
by the vacancy density $\varrho_{\rm vac}$.

\subsection{Line defects}

Classically, the Burgers vector
${\bf b}=b^i\,{\bf E}_i$
is  taken as a measure of the
dislocations in the sense that\footnote{In accord with our use of the
spatial description, the derivatives are with respect to the current position
{\bf x}.}  
\begin{equation}
b^i=   \int_{\partial{\mathcal{ C}}} {\bf e}^i\cdot d{\bf l}
=\int_{\mathcal{ C}}(\Curl\, {\bf e}^i)\!\cdot\!{\bf n}_{_\mathcal{ C}}
\,da,\label{burgers}
\end{equation}
represents the atomic lattice displacements in the reference lattice for a
circuit
$\partial{\mathcal{ C}}$ of a fixed surface
$\mathcal{C}\subset\mathcal{B}$  with unit normal ${\bf n}_{_\mathcal{ C}}$
(Bilby \& Smith, 1956; Davini, 1986). To see this, notice that the quantity $b^{i}$ 
defined by 
(\ref{burgers}) is the net number of $i$-planes (i.e., planes orthogonal
 to $\bfe^{i}$) which are crossed by the curve $\partial{\cal C}$, 
 so that any term in the combination ${\bf b}= b^{i}\bfE_{i}$ represents: 
 (the net
 number of added $i$-planes encountered upon traveling along $\partial{\cal 
 C}$) times (the reference $i$-lattice vector) = (the net Burger vector of the 
 circuit). 

By definition the  reference lattice vectors
${\bf E}_i$ are constant and dislocation free.
Define the 3 quantities
\begin{equation}
\bfg^{i}({\bf x}):=\Curl \, {\bf e}^i ({\bf x}).
\end{equation}
Commonly  ${\bf E}_{i}\otimes\bfg^{i}$ is
used as a tensorial measure of the density of {dislocations}
and corresponds to the dislocation density introduced by Nye (1953).
Since  the reference lattice vectors ${\bf E}_i$ are constant, we use
instead the 3 vectorial dislocation densities $\bfg^{i}$, which are taken as a
spatial fields on $\mathcal{B}$.

We will frequently use the equivalent fields
\begin{equation}
g^{ij}:=\frac{1}{n}\, {\bf g}^i\!\cdot{\bf e}^j,
\end{equation}
which can be interpreted as  the
\emph{dislocation density per unit cell}
 (Davini, 1986; Davini \& Parry, 1991).

We will model phase boundaries as sharp surfaces. To allow for
this possibility,
suppose now that the circuit $\partial\mathcal{C}$ is intersected by
a sharp surface $\mathcal{S}$ across which there is a
jump in lattice vectors.
Then (\ref{burgers}) must be modified to
\begin{equation}
b^i=   \int_{\partial{\mathcal{ C}}} {\bf e}^i\cdot d{\bf l}
=\int_{\mathcal{ C}}(\Curl\, {\bf e}^i)\!\cdot\!{\bf n}_{_\mathcal{ C}}\,da
- \int_{\partial{\mathcal C}\cap{\mathcal S}} \ljump {\bf e}^i\rjump \cdot
d{\bf l},
\end{equation}
where $\ljump\cdot\rjump$ denotes the jump of a quantity across the surface
$\mathcal{S}$ in which the jump is taken as the
limit from the portion of $\mathcal{B}$ into which the normal
${\bf n}_{_\mathcal{ S}}$ to the surface
points.
Since the above relation is independent of the surface
$\mathcal{ C}$ (keeping fixed the boundary curve $\partial \mathcal{C}$),
the tangential jump in the lattice vectors provides  a
measure of the defectivity of the interface. Thus we introduce
\begin{equation}
 \ljump  {\sf P}{\bf e}^i\rjump
 \label{surface_density}
\end{equation}
as 3 vectorial measures of the \emph{surface dislocation density},
 where
\begin{equation}
{\sf P}:={\bf 1}-{\bf n}_{_\mathcal{ S}}\!\otimes{\bf n}_{_\mathcal{ S}}
\end{equation}
 is the projection operator onto the tangent plane of $\cal S$.
The quantity  $\ljump {\bf E}_{i}\otimes ({\sf P}{\bf e}^i)\rjump $
corresponds to the
tensorial surface density introduced by Bilby (1955). It is equivalent to
(\ref{surface_density}) since the reference lattice vectors are continuous
and linearly independent.
To see that  (\ref{surface_density}) indeed measures the density of 
interfacial dislocations, 
let ${\bf u}$ be a tangent vector to the 
 interface $\cal S$. Then 
\begin{equation}
\ljump {\bf E}_{i}\otimes ({\sf P}{\bf e}^i)\rjump{\bf u}=
(\ljump{\bf e}^i\rjump\! \cdot \!{\bf u})\,{\bf E}_{i}. 
\label{surface_burgers}
\end{equation}
And since 
 $(\bfe^{i})^{\alpha}\!\cdot{\bf u}$ and 
$(\bfe^{i})^{\beta}\!\cdot{\bf u}$ are 
the net number of $i$-planes 
 from side $\alpha$ or $\beta$  which  
  intersect the vector $\bf u$ (recall that $|\bfe^{i}|$ is 
  the reciprocal of the distance between two consecutive $i$-planes),   
  $\ljump\bfe^{i}\rjump\cdot{\bf u}$ is the net number of $i$-planes 
 intersecting  ${\bf u}$. Thus  the sum  (\ref{surface_burgers})
consists of   the net
 number of added $i$-planes encountered upon traveling along ${\bf u}$
 times the reference $i$-lattice vectors, which equals
the net Burgers vector of the 
 interfacial  
 dislocations  along ${\bf u}$.

In classical elastostatics, a deformation is holonomic (i.e., compatible or
integrable) if the strain tensor is  the gradient of
a global, smooth map. For our model of a crystalline solid,
this notion is expressed by a 
compatibility condition stating that there exists a smooth map ${\bfchi}$
such  that
\begin{equation}
{\bf E}_i\otimes  {\bf e}^i=\Grad\, \bfchi.\label{holonomic_lattice}
\end{equation}
In other words, the state of a crystalline solid is holonomic
if the {inverse lattice distortion}
${\bf F}_{\rm lat}$ is  the gradient of
a global, smooth map. In this case, the notion of deformation arises naturally
since ${\bfchi}$ corresponds to the
\emph{macroscopic inverse deformation} to the reference lattice.
We will call a lattice a \emph{holonomic lattice} (with respect to
${\bf E}_i$) if
(\ref{holonomic_lattice}) is satisfied
for some $\bfchi$. The reference lattice, by definition,
is holonomic; the distorted lattice however is in general not holonomic.
And since the reference lattice vectors ${\bf E}_i$ are constant,
(\ref{holonomic_lattice})
shows that the bulk dislocation densities $\bfg^i$ vanish identically for a
holonomic
lattice (though the surface dislocation densities (\ref{surface_density})
need not vanish
at a singular surface $\cal S$ of a holonomic lattice).
In this model of a crystalline solid, 
dislocations are viewed as the obstruction
to patching together local measurements of the reciprocal lattice to
a single, global lattice.

\subsection{Born rule}

The \emph{Born rule} states that lattice vectors deform as material
line elements under a given deformation  (Ericksen, 1984).
We are primarily concerned  with the case when this does not
hold, as the Born rule excludes dislocations in the bulk. 
We will  consider however the consequences of imposing it
as a constraint, where it will be shown to imply common models
for elastic equilibrium as special cases. 

Consider now a holonomic lattice in the actual configuration,
so that the dislocation density vanishes inside $\mathcal{B}$.
There is  a natural  inverse deformation $\bfchi$ for the
lattice, through which
we can associate $\bfchi(\mathcal{B})$ with the region occupied in the
undeformed configuration. In this case, the Born rule yields
(cf.\ (\ref{holonomic_lattice}))
\begin{equation}
{\bf e}_{i}=(\Grad\,\bfchi^{-1}){\bf E}_{i}
\quad
\text{or}
\quad
{\bf e}^{i}=(\Grad\,\bfchi)^{\top}{\bf E}^{i}.
\label{Born rule}
\end{equation}
 %
If we assign  on $\bfchi(\mathcal{B})$ the perfect lattice
determined by ${\bf E}^{i}$, then (\ref{Born rule}a)
  states that the
lattice vectors deform as material vectors through $\bfchi^{-1}$.
Alternatively,
 (\ref{Born rule}b) states that the reciprocal lattice
vectors deform as material vectors through  $\bfchi$.

Using the Born rule at a surface of discontinuity $\mathcal{S}$,
the 3 surface dislocation densities (\ref{surface_density}) become
\begin{equation} \ljump  {\sf P}(\Grad\,\bfchi)^{\top}\rjump{\bf E}^i,
 \end{equation}
 with  ${\sf P}$  the projection operator onto the tangent plane of $\cal S$.
They correspond to the single tensorial quantity
$\ljump  {\sf P}(\Grad\,\bfchi)^{\top}\rjump$ used by 
Cermelli \& Gurtin (1994).

\section{Lattice Variations}

Let  $({\bf e}_i,\varrho)$ denote as before a state of a body
 occupying the region $\mathcal{B}$.  We denote an
 arbitrary smooth variation of this state by
$(\delta{\bf e}_i, \delta\varrho)$,
which does not necessarily correspond to any equilibrium
state.\footnote{Since we are using a spatial description,
the variations correspond to the Eulerian variations of Leo \& Sekerka
(1989)} The external boundary $\partial\mathcal{B}$ is assumed rigid
and fixed, so  the variations  vanish
on  $\partial\mathcal{B}$
(but not necessarily on an internal boundary such as the phase interface).
Consider now the additive composition of the two configurations
\begin{equation}
{{\bf e}_i}_\lambda:={\bf e}_i+\lambda\,\delta{\bf e}_i, \qquad
\varrho_\lambda:=\varrho+\lambda\, \delta\varrho
\end{equation}
where $\lambda$ is a small parameter.
This composition induces a corresponding variation
in any functional $\Phi$ depending on the composed configuration:
\begin{equation}
\delta \Phi=\frac{ \partial }{\partial \lambda}
\Phi[{{\bf e}_i}_\lambda,\varrho_\lambda]\bigg|_{\lambda=0}.  
\end{equation}
In particular, it induces  variations in dislocation density ${\bf
g}^{i}$ and the reciprocal lattice vectors  ${\bf e}^i$, related by
the identity
\begin{equation}
\delta \bfg^i=\Curl\, \delta{\bf e}^i.
 \label{flux}
\end{equation}
The induced variation of the
Burgers vector is $\delta {\bf b}=\delta b^i \,{\bf E}_i$
where
\begin{equation}
\delta {b}^i=\int_{\partial{\mathcal{ C}}}
        \delta {\bf e}^i\cdot d{\bf l}.
\label{burgersrate}
\end{equation}
Generally the  $\delta{\bf e}^i$ induce a
\emph{dislocation flux} (Kosevich, 1962).

A straightforward calculation yields the relations
\begin{align}
\delta n&=n\,{\bf e}_i\cdot\delta{\bf e}^i, \notag
\\
n \,\delta g^{ij}&=- ({\bf g}^i\!\cdot{\bf e}^j){\bf e}_l\!\cdot\delta
{\bf e}^l
+{\bf e}^j\!\cdot\delta{\bf g}^i
+{\bf g}^i\!\cdot\delta{\bf e}^j,\label{dislocation density}
\end{align}
which will be needed later.

Consider now  an internal boundary given by
the singular surface ${\cal S}$. We allow the
${\bf e}_i$ and
the dislocation density $g^{ij}$ to be discontinuous across this surface.
A variation of the position of the interface in  space is described by
\begin{equation}
\delta{\bf r}={\bf n}_{_\mathcal{ S}} \delta{ r},
\end{equation}
with ${\bf n}_{_\mathcal{ S}}$ a choice of unit normal to $\cal S$.
Variations of the position of the surface  $\cal S$ and of
 lattice vectors at the surface  will involve in general the creation
and transport of defects.

\section{Equilibrium Equations without Interfacial Energy}

In section 2 we developed the kinematics to describe an equilibrium
state of a crystalline solid with point and line defects.
In section 3 we extended the kinematic structure in order to
discuss variations of states
that induce creation and transport of such defects.
In this section we employ this kinematic structure to
derive the equations governing multi-phase equilibria.
The approach is variational and involves
rendering stationary an appropriate energy potential.
We treat crystal-crystal and crystal-melt equilibria separately.
Also we consider first the case without interfacial energy, then
in the following section extend the equilibrium results to include
interfacial energy.

\subsection{Crystal-crystal equilibrium}

Let two crystalline solid phases (denoted by $\alpha$ and $\beta$)
share a common interface
${\cal S}$ and be immersed in a heat bath at temperature $\theta_B$.
Further, let the external boundaries be impermeable and fixed.
Such a situation can be in equilibrium only if an appropriate
thermodynamic potential is stationary.
We now construct such a potential
and determine the necessary conditions for it to be stationary.
 We continue to use a spatial description and
take all extensive quantities
 describing the crystalline solid
as per unit cell in the actual state.

Since there is no external mass supply,
the total mass of the system is constant and is given by
\begin{align}
M &= M^\alpha + M^\beta
  = \int_{{\cal B}^\alpha\cup{\cal B}^\beta} n \varrho \, dv.
\label{mass}\end{align}
%
%
We assume that the total internal energy of the system is given by
\begin{align}
E &= E^\alpha + E^\beta
  = \int_{{\cal B}^\alpha\cup{\cal B}^\beta} n \epsilon\, dv,
\label{energy}\end{align}
where $\epsilon$ is the internal energy density per unit cell
in the actual state,
and any surface energy has been assumed negligible.
Furthermore, we assume that the entropy is additive:
\begin{align}
S & =S^\alpha +S^\beta
   =\int_{{\cal B}^\alpha\cup{\cal B}^\beta} n s \, dv,
 \label{entropy}\end{align}
where $s $ is the entropy density per unit cell.

A constitutive relation must specify one these quantities as a function
of the current  state of the material. As a local
measure of the lattice
structure, we
choose the quantities
$({\bf e}_i, g^{ij}, \varrho)$, since ${\bf e}_i$ specifies
the periodic structure of the lattice, $ g^{ij}$ the dislocation
density per unit cell, and $\varrho$ the atoms per unit cell.
This choice is not unique;  it is also
equivalent to $({\bf e}^i, \bfg^{i}, \varrho_{\rm vac})$.
As a measure of the local thermodynamic state, we choose the entropy $s$
per unit cell.
Our constitutive assumptions are then
that the internal energy per unit cell of
each  phase is given by:
\begin{align}
\epsilon&={\epsilon}^\alpha({\bf e}_i, g^{ij},
                 \varrho, s) \qquad \mbox{for phase $\alpha$} ,\notag\\
\epsilon&={\epsilon}^\beta({\bf e}_i, g^{ij},
                 \varrho, s)  \qquad \mbox{for phase $\beta$} .
\label{constitutive relation}
\end{align}
Notice that  we assume that the energy
depends on the current state as determined by the
\emph{local lattice structure}
and entropy. In particular, there is no dependence on a macroscopic deformation
gradient since we have not introduced a fixed reference configuration
for ${\cal B}^\alpha\cup{\cal B}^\beta$.

Following Gibbs (1878), the \emph{grand canonical potential}
\begin{equation}
 \Omega:= E -\theta_B\,  S - \mu\, M 
\label{grand-potential}
\end{equation}
(at constant total mass and
total entropy) is stationary in equilibrium, where
$\mu$ is a constant Lagrange multiplier ensuring conservation of mass.
A necessary condition for its stationarity is that the
first variation vanish:
\begin{equation}\delta \Omega=
\delta E -\theta_B\, \delta S - \mu\, \delta M  = 0.
\label{gibbs}
\end{equation}
Thus using (\ref{mass})--(\ref{gibbs})
along with (\ref{dislocation density})
 and denoting by
$$
\ljump h\rjump=h^{\beta}-h^{\alpha}
$$
the jump of a field $h$ at the interface from the
$\alpha$ and $\beta$-sides,\footnote{The choice of ${\bf n}_{_\mathcal{ S}}$
pointing outward from phase $\alpha$
is consistent with our previous convention on the
use of $\ljump\cdot\rjump$} we have
\begin{align}
\delta \Omega &=
    \delta\int_{{\cal B}^\alpha\cup{\cal B}^\beta}n \epsilon\,dv
   -\theta_B\, \delta\int_{{\cal B}^\alpha\cup{\cal B}^\beta}n s\,dv
   -\mu \,\delta \int_{{\cal B}^\alpha\cup{\cal B}^\beta} n \varrho \,
dv \notag \\
&= \
\int_{{\cal B}^\alpha\cup{\cal B}^\beta}
\big[ n (\epsilon -\theta_B s -\mu \varrho)\,{\bf
e}_i\cdot\delta{\bf e}^i+
n(\frac{\partial \epsilon}{\partial\varrho}-\mu)\,
\delta\varrho\notag \\
& \qquad \qquad \qquad
+n(\frac{\partial \epsilon}{\partial s}-\theta_B)\,
\delta s   +
n\frac{\partial \epsilon}{\partial{\bf e}_i}
\cdot\delta{\bf e}_i
+
n\frac{\partial \epsilon}{\partial g^{ij}}\,
\delta g^{ij} \big]  \,dv\notag \\
& \qquad \qquad
-\int_{\cal S}\ljump n (\epsilon-\theta_B s-\mu \varrho)\rjump\, \delta{ r}\,da
=0\label{variation-0}
\end{align}
where the surface integral arises from the variation of the interface
and reflects the induced change in phase, i.e., accretion.

The main difficulty is to choose an appropriate set of independent
variations in the bulk and at the interface
that yields equations in a physically transparent form.
To this end, let
\begin{equation}
\omega:=\epsilon-\theta_B s -\mu\varrho.
\end{equation}
denote the \emph{grand canonical potential per unit cell}.
Further we introduce
\begin{equation}
{\bf k}_i:=\frac{\partial \omega}{\partial g^{ij}}\, {\bf e}^j,
\qquad
{\bf t}^i:=n \frac{\partial \omega}{\partial{\bf e}_i},
\qquad
{\bf t}_i:=({\bf t}^j\!\cdot{\bf e}_i){\bf e}_j,
\label{defs}
\end{equation}
and  we identify
${\bf E}^i\otimes{\bf k}_i$ as the \emph{dislocation couple tensor} and
${\bf e}^i\otimes{\bf t}_i$ as the \emph{Cauchy
stress tensor} (Cermelli \& Sellers, 1998).
The choice of basis indicates that the Cauchy
stress acts on the actual lattice, whereas the dislocation couple acts
on the reference lattice.
We also introduce
the \emph{configurational stress tensor } ${\bf E}^i\otimes{\bf c}_i $
where\footnote{The  generalized Eshelby relation
(\ref{eshelby_relation}) has been derived by an invariance argument
in Cermelli \& Sellers (1998).}
\begin{equation}
{\bf c}_i := \frac{\partial( n \omega)}{\partial{\bf e}^i}=
n \omega\, {\bf e}_i-
{\bf t}_i-
({\bf g}^j\times{\bf e}_i)\times{\bf k}_j.
\label{eshelby_relation}
\end{equation}
is a generalized Eshelby relation containing an additional term
due to the dislocations.
Again the choice of basis indicates that the configurational stress,
in contrast to the Cauchy stress, acts
on the reference lattice.
In terms of the quantities (\ref{defs}) and (\ref{eshelby_relation})
the variation (\ref{variation-0}) can be expressed alternatively  as
\begin{align}
\delta \Omega &=
\int_{{\cal B}^\alpha\cup{\cal B}^\beta}
\big[
n(\frac{\partial \epsilon}{\partial\varrho}-\mu)\,
\delta\varrho +n(\frac{\partial \epsilon}{\partial s}-\theta_B)\,
\delta s   
 +
({\bf c}_i
+\mbox{\Curl\,}{\bf k}_i)
\cdot\delta{\bf e}^i
  \big]  \,dv
   \notag\\
&\qquad\qquad\qquad
-\int_{\cal S}\big( \ljump
 n\omega\rjump\, \delta r
+
{\bf n}_{_\mathcal{ S}}\!
\cdot\ljump{\delta}{\bf e}^i\times{\bf k}_i\rjump \big) \,da =0,
\label{incoherent-0}
\end{align}
where we have used the divergence theorem and
(\ref{flux}).
The expression (\ref{incoherent-0}) of the first variation
of the energy functional
is now in a convenient form for deriving the appropriate necessary
conditions.

The choice of the possible independent variations
should reflect the relevant physical situation. We consider two
different cases: one corresponding to no constraints on the possible
variations, the other corresponding to when holonomicity is imposed
as a constraint, thus excluding variations
that create dislocations
in the bulk.

\subsubsection{Case I: unconstrained variations }

\noindent{\sl $\bullet$ Bulk Conditions}: \quad We take
\begin{equation}
\delta\varrho, \qquad \delta s, \qquad
\delta{\bf e}^{i} \label{variations}
\end{equation}
as the admissible class of  independent
variations in the continuous bulk regions. These variations
allow defect creation and transport in the
bulk.
With  (\ref{incoherent-0}) they yield the following
necessary conditions: %
\\

\begin{itemize}
\item[-]\emph{uniform chemical potential}
\begin{equation}
\frac{\displaystyle\partial \epsilon^\alpha}{\displaystyle\partial\varrho}
=\frac{\displaystyle\partial \epsilon^\beta}{\displaystyle\partial\varrho}
=\mu ; \label{chemical_potential}\end{equation}
\item[-]\emph{uniform temperature}
\begin{equation}
\frac{\displaystyle \partial \epsilon^\alpha}{\displaystyle\partial s}=
\frac{\displaystyle \partial \epsilon^\beta}{\displaystyle\partial s}
=\theta_B ;\label{temperature}\end{equation}
\item[-]\emph{dislocation couple balance for each phase}
\begin{equation}
{\bf c}_i+{\Curl\,}{\bf k}_i ={\bf 0}.
\label{couple}
\end{equation} \\[-5mm]
\end{itemize}
The uniformity of the chemical potential (\ref{chemical_potential}) and
temperature (\ref{temperature}) are standard and represent chemical
and thermal equilibrium. The dislocation couple balance (\ref{couple})
represents mechanical equilibrium.
Additionally, it  implies immediately the
\emph{configurational force balance for each phase}
\begin{equation}
\Div\,{\bf c}_{i}=0,
\label{configurational}
\end{equation}
%
which, with the Eshelby relation
(\ref{eshelby_relation}),
is itself equivalent to the \emph{Cauchy force balance}
\begin{equation} {\Div}\, ({\bf e}^i\otimes{\bf t}_i) ={\bf 0} ,
\label{cauchy}
\end{equation}
which is the more standard mechanical equilibrium condition.
But provided that the  lattice variations $\delta{\bf e}^{i}$ are
arbitrary, relations (\ref{chemical_potential})--(\ref{couple})
are the  \emph{independent} necessary conditions for equilibrium in the bulk. \\

\noindent{\sl $\bullet$ Interfacial Conditions}: \quad
%
%
We want to allow for defect creation and transport at the
interface so that we take the variations  of the 
lattice vectors $\delta{\bf e}^{i}$ on each side of the phase interface
as independent, but require that $\delta{r}$ be continuous
across the interface since
the two phases remain in contact.
Thus the independent  variations at $\cal S$  are
\begin{equation}
\delta{r},\quad
({\delta}{\bf e}^i)^{\alpha}, \quad
({\delta}{\bf e}^i)^{\beta}. \label{incoherent_variables}
\end{equation}
%
They correspond to an \emph{incoherent interface}.
With this  choice of independent variations at the surface,
(\ref{incoherent-0}) yields as necessary interfacial conditions:

\begin{itemize}
\item[-]\emph{continuity of the grand canonical potential}
\begin{equation}
\ljump n\omega\rjump=0;
\label{configuration_traction}
\end{equation}
\item[-]\emph{vanishing  of the individual tangential couples}
 %
%
\begin{equation}
{\sf P}({\bf k}_i)^{\alpha}={\bf 0} 
\qquad \text{and} \qquad
{\sf P}
({\bf k}_i)^{\beta}={\bf 0} .
\label{couple_traction} 
\end{equation} 
\\[-5mm]
\end{itemize}
The continuity of the grand canonical potential (\ref{configuration_traction})
is standard and is
due to the fact that the phases may not separate at
the interface. The vanishing  of the individual tangential couples
reflects the fact that 
an incoherent interface must be in mechanical
equilibrium with respect to both
lattices.
Moreover, this latter  condition
implies in turn the 
\begin{itemize}
\item[-]\emph{vanishing of the
individual configurational tractions}
\begin{equation}
({\bf c}_i)^{\alpha}\cdot{\bf n}_{_\mathcal{ S}}=0
\qquad \text{and} \qquad
({\bf c}_i)^{\beta}\cdot{\bf n}_{_\mathcal{ S}}=0.
\label{equivalent relations-interface1}
\end{equation}
\end{itemize}
As in the corresponding bulk conditions on the configurational stress,
(\ref{equivalent relations-interface1})  are \emph{not}  independent
 conditions since  the
identity
\begin{equation}
\int_{S^\prime}({\bf c}_i)^{\alpha}\cdot{\bf n}_{_\mathcal{ S}}\,da
=-\int_{\partial S^\prime}({\bf k}_i)^{\alpha}\cdot d{\bf l},
\end{equation}
(which holds for any subsurface $\mathcal{S}^\prime\subset \mathcal{S}$)
and (\ref{couple_traction}) imply (\ref{equivalent
relations-interface1}). Additionally, the identity
\begin{equation}
{\bf n}_{_\mathcal{ S}}\! \cdot \ljump
{\bf e}^{i}\otimes{\bf t}_{i}
\rjump
{\bf n}_{_\mathcal{ S}}=\ljump n\omega\rjump - 
{\bf n}_{_\mathcal{ S}}\! \cdot \ljump
{\bf e}^{i}\otimes{\bf c}_{i}\rjump{\bf n}_{_\mathcal{ S}}
- {\bf n}_{_\mathcal{ S}}\! \cdot \ljump
{\bf e}^{i}\otimes(({\bf g}^j\times{\bf e}_{i})\times{\bf k}_j)
\rjump{\bf n}_{_\mathcal{ S}},
\end{equation}
which follows from the generalized Eshelby relation (\ref{eshelby_relation}),
shows that the continuity of the grand canonical potential
is equivalent  with (\ref{couple_traction}) and (\ref{equivalent
relations-interface1}) to the 
\begin{itemize}
\item[-]\emph{continuity of the normal
Cauchy traction}
\begin{equation}
{\bf n}_{_\mathcal{ S}}\!\cdot
\ljump
{\bf e}^{i}\otimes{\bf t}_{i}
\rjump
{\bf n}_{_\mathcal{ S}}
=0.
\label{equivalent relations-interface2}
\end{equation}
\end{itemize}
Thus  the \emph{independent} equilibrium conditions  at the incoherent interface
 are (\ref{couple_traction}) and either  (\ref{configuration_traction})
or, equivalently, (\ref{equivalent relations-interface2}). 


\subsubsection{Case II: holonomic variations (Born rule)}

Since dislocations are in general created or transported 
when the lattice variations $\delta {\bf e}_i$ are arbitrary, 
the  configurational force balance (\ref{configurational}) and
corresponding interface
conditions (\ref{equivalent relations-interface1}) are consequences of
the dislocation couple balance (\ref{couple}) and corresponding interface
condition (\ref{couple_traction}).
We now introduce
a restricted class of lattice variations that yield (\ref{configurational}) and
 (\ref{equivalent relations-interface1}) as
\emph{independent} necessary conditions.

Consider the class of holonomically constrained variations
for which  the Born rule (\ref{Born rule}) holds:
variations of the lattice basis $\delta{\bf e}^i$ are such that
$\Curl (\delta{\bf e}^i)=0$.    Thus away from the surface $\mathcal{S}$
the variations
can be expressed
in terms of the inverse  deformation $\bfchi$
by the relation
\begin{equation}
\delta{\bf e}^i=(\Grad \,\delta\bfchi)^\top{\bf E}^i.
\label{existence of the chi}
\end{equation}
Since the external boundary $\partial {\cal B}$ is fixed,
$\delta{\bfchi}$  vanishes at 
$\partial {\cal B}$, which excludes any rigid body 
displacements. {Thus the actual and 
varied states may be realized as  
deformations of the same perfect-lattice reference configuration 
$(\bfchi({\cal 
B}),{\bf E}_{i})$. In the terminology of Davini and Parry 
(1989), the actual and varied states are {\em elastically related}. 
This is false for 
arbitrary variations not vanishing at $\partial {\cal B}$, since in 
that case   rearrangements\footnote{Rearrangements are changes 
of state which do not  modify the local lattice microstructure.
A  discussion of the equilibrium conditions for a 
crystalline body when global rearrangements  are allowed, is 
due to Fonseca and Parry (1992).}
 such as slip may occur,
so that  the states ${\bf e}^{i}$ and ${\bf e}^{i}+\delta{\bf e}^{i}$ 
are only {\em neutrally related}.
} 

The Born rule excludes dislocations in the bulk but not necessarily at an
incoherent interface. Remember that
at a \emph{coherent} interface, $\bfchi$ is  continuous, whereas at
an  \emph{incoherent} interface, $\bfchi$ is  discontinuous.
Thus with the Born rule (\ref{Born rule}), 
 the continuity of $\bfchi$ implies the vanishing of the surface dislocation
density (\ref{surface_density}) whereas discontinuities in $\bfchi$ may
lead to nonvanishing surface dislocations.

In order to represent these two possibilities more succinctly,
%
we introduce the quantities
\begin{equation}
\delta_{\mathcal{S}}\chi^{i}:=\big(\delta\bfchi
+  ({\sf grad}_{{\bf n}_{\mathcal{ S}}}\bfchi)\,\delta r\big)\cdot{\bf
E}^{i},  \qquad  
{\sf grad}_{{\bf n}_{\mathcal{ S}}}\bfchi:= 
   (\Grad\bfchi){\bf n}_{_\mathcal{ S}},
\end{equation}
which are, respectively, the components of the
variation of the inverse deformation $\bfchi$ following
 the variation of the interface $\mathcal{S}$
and the derivative of the deformation normal to the current interface.  
Note that
$\delta_{\mathcal{S}}\chi^{i}$ is continuous when
$\bfchi$ is continuous:
\begin{equation}
\ljump\bfchi\rjump={\bf 0}
\quad
\Longrightarrow
\quad
\ljump\delta_{\mathcal{S}}\chi^{i}\rjump=0.
\end{equation}
%

By a procedure analogous as before, we may write the first variation
of the energy functional in the form
\begin{align}
\delta \Omega  =& 
\int_{{\cal B}^\alpha\cup{\cal B}^\beta}
\big[
n(\frac{\partial \epsilon}{\partial\varrho}-\mu)\,
\delta\varrho +n(\frac{\partial \epsilon}{\partial s}-\theta_B)\,
\delta s  
+ (\mbox{\Div\,}{\bf c}_i)\,
\delta\bfchi\cdot{\bf E}^{i}
  \big]  \,dv  \notag\\
&\ \qquad\qquad\qquad \
-\int_{\cal S}\big( \ljump
 {\bf c}_i \,\delta_{\mathcal{S}}\chi^i\rjump \cdot{\bf n}_{_\mathcal{ S}}
-
{\bf n}_{_\mathcal{ S}}\cdot\ljump
{\bf e}^i\otimes{\bf t}_i
\rjump{\bf n}_{_\mathcal{ S}}\, \delta r \big) \,da =0,
\label{incoherent}
\end{align}
where now the configurational stress component is defined as
\begin{equation}
{\bf c}_i := \frac{\partial (n\omega)}{\partial{\bf e}^i}=
n \omega\, {\bf e}_i-
{\bf t}_i,
\label{eshelby relation2}
\end{equation}
so that
\begin{equation}
(\Grad \bfchi)^{\top}({\bf E}^i\otimes{\bf c}_i) =
n \omega\, {\bf 1}-
{\bf e}^i\otimes{\bf t}_i
\label{eshelby relation3}
\end{equation}
corresponds to the usual Eshelby relation as expressed in the
spatial description. As there are now no dislocations in the bulk,
there is no dislocation contribution to (\ref{eshelby relation3}) and 
the configurational stress coincides with the classical Eshelby stress.\\

\noindent{\sl $\bullet$ Bulk Conditions}: \quad Taking as
independent variations in
the bulk

\begin{equation}
\delta\varrho,\quad
\delta s,\quad
\delta\bfchi,
\end{equation}
we obtain now the following  necessary conditions:\\

\begin{itemize}
\item[-]\emph{uniform chemical potential } (\ref{chemical_potential});

\item[-]\emph{uniform temperature } (\ref{temperature});
\item[-]\emph{configurational force balance {\rm (\ref{configurational}) }
for each phase}.\\
\end{itemize}
Because of the Eshelby relation
(\ref{eshelby relation2}), the configurational
force balance  (\ref{configurational}) in the bulk  is equivalent to
the Cauchy force balance (\ref{cauchy}).  But for this holonomic
case, the dislocation couples ${\bf k}_i$ are \emph{constitutively
indeterminate} and are now defined by the couple balance (\ref{couple}).
Thus the \emph{independent} conditions with the class of holonomic
variations are (\ref{chemical_potential}), 
(\ref{temperature}), and either (\ref{configurational}) or (\ref{cauchy}).
\\


\noindent{\sl $\bullet$ Incoherent interfacial conditions}: \quad
We want to allow for  surface dislocations at an \emph{incoherent interface},
so that we take as the admissible class of independent variations
\begin{equation}
\delta{r},\qquad
(\delta_{\mathcal{S}}\chi^i)^{\alpha},
\qquad
(\delta_{\mathcal{S}}\chi^i)^{\beta}. 
\end{equation}
With this choice of independent variations,
(\ref{incoherent}) yields the conditions:
\\

\begin{itemize}
\item[-]\emph{vanishing the individual configurational tractions}
(\ref{equivalent relations-interface1});
%
%
%
%
%
\item[-]\emph{continuity of the normal Cauchy  traction}
(\ref{equivalent relations-interface2}). \\
\end{itemize}
 %
Additionally, the continuity of the grand canonical potential 
(\ref{configuration_traction}) is
still equivalent to the continuity of the
normal Cauchy traction (\ref{equivalent relations-interface2}), by
relations (\ref{equivalent relations-interface1})
and
(\ref{eshelby relation2}). 

The relations (\ref{equivalent relations-interface1}), 
(\ref{equivalent relations-interface2}) for incoherent interfaces with
the class of holonomically constrained
variations correspond precisely
to the equilibrium results of Cermelli \& Gurtin
(1994) in the spatial description. The equivalent
relations (\ref{configuration_traction}), 
(\ref{equivalent relations-interface1})
correspond to the
 expressions derived by Larch\'e \& Cahn (1978). \\
%
%
%


\noindent{\sl $\bullet$ Coherent interfacial conditions}: \quad
There are no surface dislocations associated with a \emph{coherent} interface.
%
Furthermore, variations that keep the interface coherent should preserve the
continuity of $\bfchi$ at  $\cal S$, so that we restrict the class
of variations to those such that  %
\begin{equation}
\ljump\delta_{\mathcal{S}}\chi^i\rjump=0.
\end{equation}
Thus the independent  variations at $\cal S$  are
\begin{equation}
\delta{r},\qquad
\delta_{\mathcal{S}}\chi^i.
\end{equation}
With this choice of independent variations,
 (\ref{incoherent}) yields the
necessary conditions:
\\

 \begin{itemize}
\item[-] \emph{continuity of the configurational traction}
\begin{equation}\ljump{\bf c}_i\rjump\!\cdot\!{\bf n}_{_\mathcal{ S}}=0;
\label{configuration_traction2}
\end{equation}
\item[-]\emph{continuity of the normal Cauchy  traction}
(\ref{equivalent relations-interface2}). \\
%
%
\end{itemize}
These two conditions are equivalent to  
\\

 \begin{itemize}
\item[-]\emph{continuity of the Cauchy
traction}
\begin{equation}
\ljump{\bf e}^i\otimes{\bf t}_i
\rjump{\bf n}_{_\mathcal{ S}}={\bf 0};
\label{Cauchy_traction3}
\end{equation}
%
%
\item[-]\emph{continuity of the normal configurational
traction}\footnote{The normal configurational traction is also called the
\emph{driving traction} at the interface.} 
\begin{equation}{\bf n}_{_\mathcal{ S}}\!\cdot\!
\ljump{\bf E}^i\otimes {\bf c}_i\rjump{\bf n}_{_\mathcal{ S}}=
\ljump  {\bf c}_i\rjump\cdot {\bf n}_{_\mathcal{ S}}=0.
\label{configuration_traction3}
\end{equation}
\end{itemize}
In this case, the continuity of the grand canonical potential
does not generally hold. The \emph{independent} relations 
(\ref{configuration_traction2}) and (\ref{equivalent relations-interface2})
(or, equivalently,
 (\ref{Cauchy_traction3}) and (\ref{configuration_traction3}))
correspond to those derived by Cermelli \& Gurtin (1994) in the context
of nonlinear elasticity.
These results are also equivalent to the more standard expressions based on 
a fixed reference configuration as originally derived by Eshelby (1970)
and later by Robin (1974), Larch\'e \& Cahn (1978), and Grinfel'd (1981).


\subsection{Crystal-melt equilibrium}

We now turn to the conditions for crystal-melt phase equilibrium.
We identify the $\alpha$-phase as the crystalline solid and
denote the melt quantities with the superscript $F$.

The external boundaries are assumed impermeable and held fixed.
Thus the total mass of the system is constant:
\begin{align}
M &= M^\alpha  + M^F
  = \int_{{\cal B}^\alpha} n \varrho \, dv +
     \int_{{\cal B}^F}  \rho^{F} \, dv,
\label{mass_melt}\end{align}
with $\rho^{F}$ the density per unit \emph{volume}  of the fluid.
We assume that the total internal energy of the system is given by
\begin{align}
E &= E^\alpha + E^F
  = \int_{{\cal B}^\alpha} n \epsilon \, dv +
     \int_{{\cal B}^F}  \epsilon^F \, dv,
\label{energy_melt}\end{align}
where $\epsilon^F$ is the internal energy density per unit \emph{volume}
of fluid,
and any surface energy has been assumed negligible.
Furthermore, we assume that the entropy is additive:
\begin{align}
S & =S^\alpha +S^F
   =\int_{{\cal B}^\alpha} n s \, dv +
     \int_{{\cal B}^F}  s^F \, dv,
\label{entropy2}\end{align}
where $s^F$ is the entropy density per unit \emph{volume} of the fluid.
Our constitutive assumptions are that:
\begin{align}
\epsilon&={\epsilon}^\alpha({\bf e}_i, g^{ij}, s,
                 \varrho), \qquad
\epsilon^F={\epsilon}^F(s^F,
                 \rho^F ).
\label{constitutive_relation_melt}
\end{align}
As before, our condition for equilibrium is
\begin{equation}
\delta\Omega=\delta E -\theta_B\, \delta S - \mu\, \delta M  = 0.
\label{gibbs2}\end{equation}

\subsubsection{Unconstrained variations}

\noindent{\sl $\bullet$ Bulk conditions}: \quad
We take
\begin{equation}
\delta\rho^F ,\qquad\delta s^F, \qquad
\delta\varrho, \qquad  \delta s, \qquad
\delta{\bf e}^{i}
\end{equation}
as the independent
variations in the corresponding continuous bulk regions.
With this  choice of independent variations,
the analog of (\ref{incoherent}) yields the following
independent conditions:
\\

\begin{itemize}
\item[-]\emph{uniform chemical potential}
\begin{equation}
n\frac{\displaystyle\partial \epsilon^\alpha}{\displaystyle\partial\varrho}
=\frac{\displaystyle\partial \epsilon^F}{\displaystyle\partial\rho}
=\mu ;\label{melt_chem} \end{equation}
\item[-]\emph{uniform temperature}
\begin{equation}
n\frac{\displaystyle \partial \epsilon^\alpha}{\displaystyle\partial s}=
\frac{\displaystyle \partial \epsilon^F}{\displaystyle\partial s}
=\theta_B ;\label{melt_temp}\end{equation}
\item[-]\emph{dislocation couple balance in the crystalline solid phase}
\begin{equation}
{\bf c}_i+\mbox{\Curl\,}{\bf k}_i ={\bf 0}. \label{couple_melt}\end{equation}
 \\[-5mm]
\end{itemize}
As before, the couple balance (\ref{couple_melt}) implies immediately the
\emph{configurational force balance in the crystalline solid phase}
%
%
which itself is equivalent to the 
\emph{Cauchy force balance in the crystalline solid
phase}. But if we restrict the admissible lattice
variations to the holonomically constrained  class,
then configurational force balance or Cauchy force balance
 replaces the couple balance (\ref{couple_melt})
as the mechanical equilibrium condition.
\\

\noindent{\sl $\bullet$ Interfacial conditions}: \quad
We choose
\begin{equation} \delta r, \qquad \delta {\bf e}_i
\end{equation}
as the independent variations at the interface $\mathcal{S}$.
They allow the creation and transport of defects in the crystalline
solid phase.
We  obtain: \\

\begin{itemize}
\item[-]\emph{continuity of the grand canonical potential}
\begin{equation}n\omega=\omega^F; \label{cont_grand_pot}
\end{equation}
\item[-]\emph{vanishing  of the tangential couples  of the
crystalline solid}
\begin{equation} {\sf P}{\bf k}_i={\bf 0}.
\label{alpha-couple}
 \end{equation} \\[-5mm]
\end{itemize}
Additionally we obtain as a consequence of (\ref{alpha-couple})
the \emph{vanishing of the configurational traction
 of the
crystalline solid}
\begin{equation}
{\bf c}_i\cdot{\bf n}_{_\mathcal{ S}}=0. \label{melt_conf_traction}
\end{equation} %
The
\emph{continuity of the normal Cauchy  traction}
\begin{equation}
{\bf n}_{_\mathcal{ S}}\cdot({\bf e}^i\otimes{\bf t}_i)
{\bf n}_{_\mathcal{ S}}=\omega^F=\epsilon^F-\theta_B s^F-\mu
\varrho^F, \label{melt_cauchy_traction}
\end{equation} %
is equivalent to (\ref{cont_grand_pot})
 at $\mathcal{S}$  provided
(\ref{alpha-couple}) 
 and the Eshelby relation  (\ref{eshelby_relation})
for ${\bf c}_i$. 

Again, if we restrict the admissible lattice variations to
the holonomically constrained class, then
(\ref{melt_conf_traction}) replaces
 (\ref{alpha-couple}) as the equilibrium conditions
at $\mathcal{S}$. In this case, 
the relations (\ref{cont_grand_pot}), (\ref{melt_conf_traction})
correspond to those derived by Larch\'e \& Cahn (1978).

\section{Equilibrium Equations with Interfacial Energy}

The presence of dislocations at the interface as well as the crystalline 
structure of the material may modify the equilibrium conditions at the interface. 
One way to take this into account is to 
introduce an interfacial energy density that depends upon the local
state of the interface.
We now extend the previous results to include such an interfacial
energy density. Since the conditions for the bulk region do not change,
we consider only the appropriate interfacial conditions.

\subsection{The interfacial energy density}

One possible choice of variables to describe the local state of the interface
is
\begin{equation}
(\bfe^i)^{\alpha}, \qquad(\bfe^i)^{\beta},\qquad \bfn_{_{\cal S}} ,
\label{interfacial_variables}
\end{equation}
which reflect  the local lattice structure on either side of the 
interface and the orientation of the interface with respect to the 
current lattice. When the Born rule holds, this choice becomes
\begin{equation}
(\Grad\bfchi)^{\alpha}, \qquad(\Grad\bfchi)^{\beta},\qquad \bfn_{_{\cal S}} ,
\label{interfacial_variables2}
\end{equation}
which corresponds to the variables used by Cermelli \& Gurtin 
(1994). Unfortunately,
 the dependence on the actual orientation is misleading, since $\bfn_{_{\cal 
 S}}$ does not properly account for  the crystallographic
orientations of the individual lattices relative to the interface.
A  choice of variables that better captures the relevant 
physics at the interface is 
\begin{equation}
{\sf P}\ljump\bfe^i\rjump,\qquad \bfn^{\alpha},\qquad \bfn^{\beta}, 
\label{interfacial_variables3}
\end{equation}
where 
\begin{equation}
\bfn^{\alpha}= \frac{ \bfn_{_{\cal S}}\cdot(\bfe_{i})^{\alpha}\,\bfE^{i}}{
  | \bfn_{_{\cal S}}\cdot(\bfe_{j})^{\alpha}\,\bfE^{j} |   }       ,
\qquad
\bfn^{\beta}=\frac{\bfn_{_{\cal S}}\cdot(\bfe_{i})^{\beta}\,\bfE^{i}}{
 |\bfn_{_{\cal S}}\cdot(\bfe_{j})^{\beta}\,\bfE^{j}|}.
\end{equation}
The choice of variables (\ref{interfacial_variables3}) is motivated 
by the observation that 
\begin{itemize}
\item
${\sf P}\ljump\bfe^i\rjump$ is the dislocation content of  the 
interface (cf.\ (\ref{surface_density}));

\item
$\bfn^{\alpha}$ and $\bfn^{\beta}$ are the unit normals to the 
interface in the reference lattices of each phase.
While it is 
standard in theories of coherent interfaces  to include  the 
contribution of the orientation of the interface normal in the 
superficial energy, for incoherent 
interfaces the energy depends on the orientation of the interface 
normal with respect to \emph{both} lattices; 

\item the fields 
$({\sf P}\ljump\bfe^i\rjump,\bfn^{\alpha},\bfn^{\beta}) $ 
on $\cal S$ may be assigned  independently from each other. 

\end{itemize}
Again, the Born rule simplifies (\ref{interfacial_variables3})  to
\begin{equation}
{\sf P}\ljump\Grad\bfchi\rjump,\qquad \bfn^{\alpha},\qquad \bfn^{\beta}, 
\label{interfacial_variables4}
\end{equation}
where now
\begin{equation}
\bfn^{\alpha}= \frac{\left((\Grad\bfchi)^{-\top}\right)^\alpha\bfn_{_{\cal S}} }{
  |\left((\Grad\bfchi)^{-\top}\right)^\alpha\bfn_{_{\cal S}} |   }       ,
\qquad
\bfn^{\beta}=\frac{\left((\Grad\bfchi)^{-\top}\right)^\beta\bfn_{_{\cal S}} }{
 |\left((\Grad\bfchi)^{-\top}\right)^\beta\bfn_{_{\cal S}} |}.
\end{equation}
The variables (\ref{interfacial_variables4}) correspond to a special case of those
used by Leo and  Sekerka (1989).

Henceforth, we shall restrict the energy density function to have the special form
\begin{eqnarray}
\varphi=\tilde\varphi({\sf P}\ljump\bfe^i\rjump,\bfn^{\alpha},\bfn^{\beta}).
\label{the interfacial energy}
\end{eqnarray}
For calculations, however, we will find it more convenient to use the more general
energy 
\begin{eqnarray}
\varphi=\hat\varphi((\bfe^i)^{\alpha},(\bfe^i)^{\beta},\bfn_{_{\cal S}}) 
\label{raw interfacial energy}
\end{eqnarray}
and then insert (\ref{the interfacial energy}) at the end of the calculation.
For example, we denote by 
\begin{equation}
\kk_{i}^{\alpha}=\frac{\partial\hat\varphi}{\partial(\bfe^{i})^{\alpha}}
\qquad
\kk_{i}^{\beta}=
 \frac{\partial\hat\varphi}{\partial(\bfe^{i})^{\beta}}
\qquad
\boldsymbol{\xi}=\frac{\partial\hat\varphi}{\partial{\bf n}_{_{\cal S}}},
\label{hat derivatives}
\end{equation}
the derivatives of (\ref{raw interfacial 
energy}) with respect to its arguments, which are essentially surface 
stresses.  They must however satisfy the following relations obtained by
the chain rule and (\ref{the interfacial energy}):
\begin{align}
\kk_{i}^{\alpha}
&=
-{\sf P}\frac{\partial\tilde\varphi}{\partial({\sf 
P}\ljump\bfe^{i}\rjump)}
-  \frac{ \bfn_{_{\cal S}}\cdot(\bfe_i)^{\alpha}}
{|{\bf n}_{_{\cal S}}\cdot({\bf e}_j)^{\alpha}\,{\bf E}^j|}
 \left({\sf P}^{\alpha}
 \frac{\partial\tilde\varphi}{\partial\bfn^{\alpha}}
 \cdot\bfE^k\right)
 (\bfe_k)^{\alpha}, \notag
 \\ 
\kk_{i}^{\beta}
&=
 {\sf P}\frac{\partial\tilde\varphi}{\partial({\sf 
 P}\ljump\bfe^{i}\rjump)}
 -\frac{ \bfn_{_{\cal S}}\cdot(\bfe_i)^{\beta}}
 {|{\bf n}_{_{\cal S}}\cdot({\bf e}_j)^{\beta}\,{\bf E}^j|}
 \left({\sf P}^{\beta}
 \frac{\partial\tilde\varphi}{\partial\bfn^{\beta}}
 \cdot\bfE^k\right)
 (\bfe_k)^{\beta},
  \notag \\
\boldsymbol{\xi}&=
  \frac{1}{|{\bf n}_{_{\cal S}}\cdot({\bf e}_j)^{\alpha}\,{\bf E}^j|}
 \left({\sf P}^{\alpha}
 \frac{\partial\tilde\varphi}{\partial\bfn^{\alpha}}
 \cdot\bfE^k\right)
 (\bfe_k)^{\alpha}
 +\frac{1}{|{\bf n}_{_{\cal S}}\cdot({\bf e}_j)^{\beta}\,{\bf E}^j|}
 \left({\sf P}^{\beta}
 \frac{\partial\tilde\varphi}{\partial\bfn^{\beta}}
 \cdot\bfE^k\right)
 (\bfe_k)^{\beta}
 \notag \\
 & \qquad
  -\bigl(\bfn_{_{\cal S}}\cdot\ljump\bfe^{i}\rjump\bigr)
  {\sf P}\frac{\partial\tilde\varphi}{\partial({\sf 
 P}\ljump\bfe^{i}\rjump)}
 -\left(
 \frac{\partial\tilde\varphi}{\partial({\sf 
 P}\ljump\bfe^{i}\rjump)}\cdot\bfn_{_{\cal S}} \right)
 {\sf 
 P}\ljump\bfe^{i}\rjump, 
 \label{derivatives relations}
 \end{align}
with projections
\begin{equation}
{\sf P}^{\alpha}:={\bf 1}-\bfn^{\alpha}\otimes \bfn^{\alpha},
\qquad 
 {\sf P}^{\beta}:={\bf 1}-\bfn^{\beta}\otimes \bfn^{\beta}.
\end{equation}
An immediate consequence of (\ref{derivatives relations}) is that the 
surface stresses $\kk_{i}^{\alpha,\beta}$ are tangential to $\cal S$:
\begin{equation}
 \kk_{i}^{\alpha}\cdot\bfn_{_{\cal S}}
  = \kk_{i}^{\beta}\cdot\bfn_{_{\cal S}}=0.
\label{restrictions on derivatives}
\end{equation}

\subsection{The variational calculation}

 When interfacial structure is taken into account, the grand 
 canonical potential (\ref{grand-potential}) includes a contribution 
 from surface energy:
 \begin{equation}
 \Omega:= E + I -\theta_B\,  S - \mu\, M 
 \qquad
 \mbox{with}\qquad 
 I=\int_{\cal S}\varphi \,\, da.
\label{grand-canonical-interface}
 \end{equation}
%
The variation of the 
interfacial energy is given by
\begin{equation}
\delta \int_{\cal S}\varphi\,da
=
\int_{\cal S}
\left\{
\kk_{i}^{\alpha}\cdot
(\delta_{\cal S}\bfe^i)^{\alpha}+
\kk_{i}^{\beta}\cdot
(\delta_{\cal S}\bfe^i)^{\beta}
+
\left({\sf div}_{\cal S}\,
\boldsymbol{\xi} - K\varphi
\right)\delta r
\right\}\,da,
\label{variation of the raw interfacial energy}
\end{equation}
where  $\delta_{\cal S}$ is the variation following the interface (i.e.,
 $\delta_{\cal S} f= \delta f+ {\sf grad}_{\bfn_{\cal S}} f=
 \delta f+ ({\sf grad} f){\bfn_{_{\cal S}}} $
for any bulk function 
 restricted to $\cal S$), 
$K$ is the mean curvature of $\cal S$,  ${\sf grad}_{\cal S}$ and ${\sf div}_{\cal 
S}$  are the 
superficial gradient and divergence  on $\cal S$, and we have used the 
identities (cf.\ Leo and Sekerka, 1989)
\begin{equation}
\delta \int_{\cal S}\varphi\,da=
\int_{\cal S}(\delta_{\mathcal{S}}\varphi - K\varphi\, \delta r)\,da,
\qquad
\delta_{\cal S}\bfn_{_{\cal S}}=-{\sf grad}_{\cal S}\,\delta r.
\end{equation}
Remember that the interfacial stresses $\kk_{i}^\alpha$,  
$\kk_{i}^\beta$ and $\boldsymbol{\xi}$
 are required to satisfy
 (\ref{derivatives relations}).

With  (\ref{variation of the raw interfacial 
 energy}) added to the surface integral in (\ref{incoherent-0}), the 
 interfacial 
 terms in the variation of the grand canonical potential 
 (\ref{grand-canonical-interface}) can be expressed as
 \begin{multline}
 \int_{\cal S}\Big\{  \left({\sf div}_{\cal S}\,
\boldsymbol{\xi} - K\varphi -\ljump
 n\omega\rjump
 -\ljump\bfn_{_\mathcal{S}}\times{\bf k}_i
\cdot{\sf grad}_{\bfn_{\cal S}}\, \bfe^i\rjump
\right)\delta r 
\cr
 +
\kk_{i}^{\alpha}\cdot
(\delta_{\cal S}\bfe^i)^{\alpha}+
\kk_{i}^{\beta}\cdot
(\delta_{\cal S}\bfe^i)^{\beta}
 -\ljump
 {\bf n}_{_\mathcal{S}}\times{\bf k}_i
 \cdot
 (\delta_{\cal S}\bfe^i)\rjump
\Big\}
\,da.
 \end{multline}

\subsubsection{Case I: unconstrained variations}

When variations are unconstrained in the sense of Section 4.1.1,
dislocations are allowed to be created and to move both in the bulk and at 
the interface. We choose as the admissible class of independent variations
at such an \emph{incoherent interface} 
\begin{equation} \delta r, \qquad (\delta_{\cal S}{\bf e}^i)^\alpha,
\qquad(\delta_{\cal S}{\bf e}^i)^\beta.
\end{equation}
This set is more convenient than (\ref{incoherent_variables})  due to the additional
interfacial structure.
 The interfacial equilibrium conditions which replace 
(\ref{configuration_traction}) and (\ref{couple_traction}) 
are thus
\begin{itemize}
\item[-]\emph{interfacial normal force balance}
\begin{equation}
{\sf div}_{\cal S}\,
\boldsymbol{\xi} - K\varphi -\ljump
 n\omega\rjump
 +\kk_{i}^{\alpha}
\cdot
 ({\sf grad}_{\bfn_{\cal S}}\bfe^i)^{\alpha}
  +\kk_{i}^{\beta}\cdot
 ({\sf grad}_{\bfn_{\cal S}}\bfe^i)^{\beta}
 =0;
\label{orientational balance}
\end{equation}
\item[-]\emph{interfacial couple balances}
\begin{equation}
\kk_{i}^{\alpha}-
 {\bf n}_{_\mathcal{S}}\times({\bf k}_i)^{\alpha}=0
\qquad \text{and} \qquad
\kk_{i}^{\beta}+
 {\bf n}_{_\mathcal{S}}\times({\bf k}_i)^{\beta}=0 .
\label{interfacial couple balance} 
\end{equation} 
%
\end{itemize}
The interfacial relations 
(\ref {interfacial couple balance})   
state that the couples 
${\bf n}_{_\mathcal{S}}\times({\bf k}_i)^{\alpha,\beta}$ acting on interfacial 
dislocations and due to the short-range interactions with 
bulk dislocations must balance pointwise  the forces 
$\kk_{i}^{\alpha,\beta}$ in the 
interface due to interfacial dislocations. 

The surface divergence of (\ref{interfacial couple balance}) along
with the symmetry of the curvature tensor imply the
\begin{itemize}
\item[-]\emph{interfacial configurational force balances}
\begin{equation}
{\sf div}_{\cal S}\, \kk_{i}^{\alpha}
-({\bf c}_{i})^{\alpha} \cdot\bfn_{\cal S}
=0,
\qquad
\mbox{and}
\qquad
{\sf div}_{\cal S}\, \kk_{i}^{\beta}
+
({\bf c}_{i})^{\beta}\cdot\bfn_{\cal S}
=0,
\label{configurational balance holo}
\end{equation}
\end{itemize}
which generalizes (\ref{equivalent relations-interface1}).
However these are not independent conditions, but follow as simple
consequences of the interfacial
couple balances (\ref{interfacial couple balance}).
Relations (\ref{configurational balance holo}) are force balances that
state that  the ``elastic'' 
configurational traction acting on any portion $\cal R$ of the interface from each phase 
 must be balanced by the  the 
interfacial contact force,  
which tends to reorganize interfacial dislocations in $\cal R$ according to their 
mutual interaction.


\subsubsection{Case II: holonomic variations.}

In this case the variations $\delta_{\cal S}{\bf e}^{i}$ 
may be expressed in terms of the inverse 
deformation through  
(\ref{existence of the chi}) and, by the identity
$$
{\sf P}\,\delta_{\cal S}\bfe^{i}=
{\sf grad}_{\cal S}(\delta_{\cal S}\chi^{i})-({\sf grad}_{\cal S}
(\bfn_{_{\cal 
S}}\, \delta r))^{\top}\bfe^{i},
$$
the surface integral in the variation of the grand canonical 
potential (\ref{grand-canonical-interface})
becomes,  granted (\ref {restrictions 
on derivatives}),
\begin{align}
-\int_{\cal S} &
\Big\{
\big[
{\sf div}_{\cal S}\,\kk_{i}^{\alpha}
-({\bf c}_{i})^{\alpha}\cdot\bfn_{_{\cal S}}
\big]
(\delta_{\cal S}\chi^{i})^{\alpha}
+
\big[
{\sf div}_{\cal S}\,\kk_{i}^{\beta}
+
({\bf c}_{i})^{\beta}\cdot\bfn_{_{\cal S}}
\big]
(\delta_{\cal S}\chi^{i})^\beta
\cr
&
-\Big( {\sf div}_{\cal S}\,
\boldsymbol{\xi} - K\varphi 
+(\bfe^{i})^{\alpha}\cdot\bfn_{_{\cal S}}
{\sf div}_{\cal S}\,\kk_{i}^{\alpha}
+(\bfe^{i})^{\beta}\cdot\bfn_{_{\cal S}}
{\sf div}_{\cal S}\,\kk_{i}^{\beta}
\cr
& \quad\qquad 
+\kk_{i}^{\alpha}
\cdot({\sf grad}_{\bfn_{\cal S}}\bfe^{i})^{\alpha}
+\kk_{i}^{\beta}
\cdot({\sf grad}_{\bfn_{\cal S}}\bfe^{i})^{\beta}
-
{\bf n}_{_\mathcal{ S}}\cdot\ljump
{\bf e}^i\otimes{\bf t}_i
\rjump{\bf n}_{_\mathcal{ S}}\Big) \,\delta r \Big\}
\,da,
\end{align}
where ${\bf c}_{i}$ is now given by (\ref{eshelby relation2}).\\


\noindent{\sl $\bullet$ Incoherent interfacial conditions}: \quad We take
\begin{equation}
\delta r, \qquad (\delta_{\cal S} \chi^i)^\alpha, \qquad (\delta_{\cal S}\chi^i)^\beta
\end{equation}
as the admissible class of independent variations.
This yields as generalizations of 
(\ref{equivalent relations-interface1}) and (\ref{equivalent relations-interface2})
the interfacial equilibrium conditions:
\begin{itemize}

\item[-]\emph{interfacial normal force balance}
\begin{multline}
{\sf div}_{\cal S}\,
\boldsymbol{\xi} - K\varphi 
+(\bfe^{i})^{\alpha}\cdot\bfn_{\cal S}\, 
{\sf div}_{\cal S}\,\kk_{i}^{\alpha}
+(\bfe^{i})^{\beta}\cdot\bfn_{\cal S}\,
{\sf div}_{\cal S}\,\kk_{i}^{\beta}
\cr
+\kk_{i}^{\alpha}
\cdot({\sf grad}_{\bfn_{\cal S}}\bfe^{i})^{\alpha}
+(\kk_{i})^{\beta}
\cdot({\sf grad}_{\bfn_{\cal S}} \bfe^{i})^{\beta}
=
{\bf n}_{_\mathcal{ S}}\cdot\ljump
{\bf e}^i\otimes{\bf t}_i
\rjump{\bf n}_{_\mathcal{ S}}
\label{orientational balance holo}
\end{multline}
\item[-]\emph{interfacial configurational force balances} 
(\ref{configurational balance holo}).
\end{itemize}
Now however (\ref{configurational balance holo}) are \emph{independent} conditions.
Furthermore, (\ref{orientational balance holo}) is equivalent, granted 
(\ref{configurational balance holo}), to 
(\ref{orientational balance}).
The equilibrium interfacial conditions  (\ref {orientational balance}) 
and (\ref{configurational balance holo}) are appropriate to a 
constrained situation in which  dislocations cannot be created 
or move in the bulk and, for this case, are equivalent to the relations derived by
 Cermelli \& Gurtin (1994).

The non-holonomic 
equilibrium conditions (\ref{interfacial couple balance}) 
are stricter than their holonomic counterparts 
(\ref{configurational balance holo}); 
interfaces which are in equilibrium governed by  (\ref{configurational 
balance holo}), i.e., when dislocations are absent in bulk, need
not  be  in equilibrium if dislocations were allowed, since these 
dislocations could exert forces on the interface. 
As an example, consider    
a grain boundary between undeformed grains in a polycrystalline 
material, and assume that the grains are
rotated one relative to another:
\begin{equation} 
 (\bfe_{i})^{\alpha} = \bfE_{i}, \qquad (\bfe_{i})^{\beta}={\bf Q}\bfE_{i},
\end{equation}
 with ${\bf Q}$ a rigid body rotation. 
Assume that dislocations are not allowed in bulk (holonomy), fix 
$\varrho=\varrho_{0}$ and $s=s_{0}$, and take 
the grand canonical potential such that, by objectivity, 
 $\omega(\bfE_{i},\varrho_{0},s_{0})=\omega({\bf 
 Q}\bfE_{i},\varrho_{0},s_{0})=0$ is a minimum, so that 
  $({\bf c}_{i})^{\alpha}=({\bf c}_{i})^{\beta}=0$ (cf.\ 
  (\ref{eshelby relation2})). Then, according to the holonomic 
  equilibrium condition (\ref{configurational balance holo}), any 
  interface such that 
  $\kk_{i}^{\alpha,\beta}$ is 
  a constant (for instance planar interfaces with a uniform 
  dislocation density) is in equilibrium. 
  On the other hand, if 
  dislocations are free to move to and from the interface from the 
  bulk, equilibrium is 
  established by the non-holonomic relations 
  (\ref{interfacial couple balance}), and  
    such an interface 
  (planar  with a uniform 
  dislocation density) may well not be at equilibrium, as for 
  instance when    
  $\kk_{i}^{\alpha,\beta}$ is 
  non-zero but ${\bf k}_{i}$ is zero. 
  \\

\subsection{The equilibrium conditions projected on a slip system}

The transport of defects is commonly interpreted in terms of a crystallographic slip 
system, where slip typically takes place along lattice vectors.
Some insight into the structure of the bulk and interfacial 
equilibrium conditions may be gained by introducing such a crystallographic slip 
system and 
projecting the equations  onto this slip system.
The resulting relations involve forces acting 
within the slip plane, which may be interpreted as forces on the
 dislocations  moving by glide on such a  plane. 

More precisely, let 
\begin{equation}
\boldsymbol{\sigma}\otimes\boldsymbol{\mu},
\end{equation}
be a  slip system, with $\boldsymbol{\sigma}$ a slip vector and 
$\boldsymbol{\mu}$ the slip-plane normal, with the restriction that 
$\boldsymbol{\mu}\cdot \boldsymbol{\sigma}=0$. 
The slip vectors are taken to be lattice vectors, which correspond to 
the Burgers vectors of the dislocations whose motion generates the 
slip, while the slip-plane normals are reciprocal lattice vectors. 

The slip system is 
measured in the reference lattice. Its counterpart in   the
current lattice is represented by the set of vectors
\begin{equation}
\sigma^{i}\,{\bf m},
\qquad
\text{with}\quad \sigma^{i}:=\boldsymbol{\sigma}\cdot{\bf E}^{i},
\label{slip-system}
\end{equation}
and
\begin{equation}
{\bf m}=\mu_{j} \bfe^{j}, 
\qquad \text{with}\quad \mu_{j}:=\boldsymbol{\mu}\cdot\bfE_{j}.
\end{equation}
Note that $\sigma^{i}$ 
and $\mu_{i}$ are constants.

Projecting the  bulk
dislocation couple balance  (\ref{couple}) onto the slip 
system (\ref{slip-system}), we obtain 
\begin{equation}
\sigma^{i}\,{\bf c}_{i}\cdot{\bf m}
+\sigma^{i}\mu_{j}\,(\bfk_{i}\cdot\bfg^{j})
+\sigma^{i}\,\Div\,(\bfk_{i}\times{\bf m})=0,
\label{bulk balance projected}
\end{equation}
with the configurational stress given by (\ref{eshelby_relation}).
Each term in (\ref{bulk balance projected}) has a suggestive 
interpretation, namely:
\begin{itemize}

\item
the term 
$\sigma^{i}\,{\bf c}_{i}\cdot{\bf m}$ is the \emph{configurational shear}  
across the slip plane (recall in fact that ${\bf 
m}$ is the current slip-plane normal, while $\boldsymbol{\sigma}$ is contained in 
the referential slip plane), and may be viewed as forcing or inhibiting 
slip along the corresponding system. It may be identified with the so-called long-range 
stress on dislocations in a dislocated crystal. 

\item 
the term 
$\sigma^{i}\, \bfk_{i}\times{\bf m}$ is the couple acting \emph{in 
the slip plane} on dislocations with Burgers vectors parallel to 
$\boldsymbol{\sigma}$. It may be identified with 
the short-range contact couple  due to the presence of other dislocations.

\end{itemize}

Projection of the interfacial couple balances (\ref{interfacial couple 
balance}) onto a phase-$\alpha$ slip system yields
\begin{equation}
(\sigma^{i})^{\alpha}\,{\bf m}\cdot
\kk_{i}^{\alpha}-
(\sigma^{i})^{\alpha} \,\bigl(({\bf k}_i)^{\alpha}\times{\bf m}\bigr)\cdot\bfn_{_{\cal S}}=0.
\label{interfacial balance projected}
\end{equation} 
%
The couple ${\bf m}\times({\bf 
k}_i)^{\alpha}$ is tangential to the slip plane, while $\kk_{i}^{\alpha}$ is
tangential to the interface. Consider now the  
intersection line (say $L$) of the slip plane with the tangent plane to the 
interface.  A straightforward calculation shows that

\begin{itemize}

\item
the term 
$(\sigma^{i})^{\alpha}\,{\bf m}\cdot\kk_{i}^{\alpha}$ represents the projection 
onto the direction ${\bf m}-({\bf m}\cdot\bfn_{_{\cal 
S}})\bfn_{_{\cal S}}$, the \emph{perpendicular to $L$ in the tangent plane to 
the interface}, of the 
interfacial force on a dislocation with Burgers vector
$\boldsymbol{\sigma}$.

\item
the term 
$(\sigma^{i})^{\alpha}\, \bigl({\bf m}\times({\bf k}_i)^{\alpha}\bigr)
\cdot\bfn_{_{\cal 
S}}$ represents the projection onto the direction $\bfn_{_{\cal S}}-({\bf 
m}\cdot\bfn_{_{\cal S}}){\bf m}$, the \emph{perpendicular to $L$ 
in the slip plane}, of the 
 couple due to bulk dislocations on a defect with Burgers vector
$\boldsymbol{\sigma}$.

\end{itemize}
A dislocation lying along the intersection of the slip 
plane and the interface  is parallel to $L$;  the only 
relevant component of any force acting on this dislocation is the 
component orthogonal to $L$.
The couple balance   
requires that these components of the opposing couples on the 
dislocation be equal. 
Therefore
(\ref{interfacial balance 
projected}) represents a  balance of  couples 
which tend to redistribute the 
dislocation density at the interface and couples which tend to move 
the dislocations away from the interface along their slip planes, due to 
the presence of bulk dislocations. 
Two such independent balances are needed, one for each 
phase, since 
dislocations may be transferred to and from the interface from each 
crystal phase, and their slip systems   are independent.

\section{Incremental Deformations (without Interfacial Energy)}

In many models of defective materials, it is common
to introduce an elastic-plastic decomposition of the macroscopic
deformation.
In our formulation there is no need to introduce the
notion of deformation from a fixed reference configuration.
In order to compare the two approaches, however, we now introduce such
an elastic-plastic decomposition. We follow the procedure of
Davini (1989) by  superposing
a conventional elastic deformation onto a dislocated crystalline solid,
where the elastic deformation does not change the defectivity of the crystal.
We then derive the necessary  equilibrium conditions for the resulting
incremental theory.

We identify $\mathcal{B}$, which corresponds to the region occupied
by the crystalline solid in a dislocated state, to  the intermediate
configuration in an elastic-plastic decomposition.
In contrast to the standard formulations of the elastic-plastic decomposition,
we do not assume that $\mathcal{B}$ is stress-free.
Now consider a  deformation ${\bf f}:\mathcal{B}\rightarrow\rr^3$
from this dislocated state that  is continuous but
possibly non-smooth
across the surface $\cal S$. Although the incremental deformation is
continuous everywhere, the interface can be incoherent due to the
presence of surface dislocations in the dislocated intermediate configuration.
We assume that
the lattice vectors transform according to the Born rule under the
deformation ${\bf f}$:
\begin{equation}
{\bf e}_{i}\mapsto{\bf\overline{e}}_{i}:={\bf F}{\bf e}_{i},
\qquad
{\bf e}^{i}\mapsto{\bf\overline{e}}^{i}:={\bf F}^{-\top}{\bf e}^{i},
\label{lat_trans}
\end{equation}
where the ${\bf\overline{e}}_{i}$ are the current lattice vectors in the
deformed region ${\bf f}(\mathcal{B})$ and ${\bf F}=\Grad\,{\bf f}$.
The dislocation density associated to the
field ${\bf\overline{e}}^{i}$ is given by
\begin{equation}
{\bf \overline{g}}^i=\Det ({\bf F}^{-1})\,
{\bf F}{\bf g}^i,\label{disl_trans}
 \end{equation}
which yields
\begin{equation}
\overline{g}^{ij}=g^{ij}.\label{disl_unitcell}
\end{equation}
Thus the dislocation density per unit cell is unaffected by the incremental
elastic deformation.

 The lattice variations of this resulting state can be expressed as
$(\delta{\bf \overline{e}}_i, \delta\varrho)$,
which, as before,  are assumed to vanish
on the external boundary $\partial\mathcal{B}$.
Consider now the additive composition of the two configurations
\begin{equation}
{\overline{{\bf e}}_i}_\lambda:={\bf \overline{e}}_i
+\lambda\,\delta{\bf \overline{e}}_i,
\qquad
{\varrho}_\lambda:={\varrho}
+\lambda\, \delta{\varrho}
\end{equation}
where $\lambda$ is a small parameter.
This composition induces a corresponding variation
in any functional $\Phi$ depending on the composed configuration:
\begin{equation}
\delta \Phi=\frac{ \partial }{\partial \lambda}
\Phi[{{\bf \overline{e}}_i}_\lambda,{\varrho}_\lambda]
\bigg|_{\lambda=0}.
\end{equation}
Note however that
\begin{equation}
\delta{\bf \overline{e}}_i=
(\delta{\bf F}){\bf e}_i
-({\bf e}_i\cdot \delta{\bf e}^j){\bf F}{\bf e}_j.
\end{equation}

Because of (\ref{lat_trans}) and (\ref{disl_unitcell}),
 we may assume that the  energy densities
are functions of the form
\begin{align}
\epsilon&={\epsilon}^\alpha({\bf\overline{e}}_i ({\bf e}_i,{\bf F}), g^{ij},
                 \varrho, s)
\qquad \mbox{for phase $\alpha$} ,
\notag\\
\epsilon&={\epsilon}^\beta({\bf\overline{e}}_i ({\bf e}_i,{\bf F}), g^{ij},
                 \varrho, s)  \qquad \mbox{for phase $\beta$},\label{epsilon}
\end{align}
%
where the unbarred quantities refer to the dislocated 
configuration $\mathcal{B}$.
%
We introduce the  \emph{incremental Piola stress}
\begin{equation}
{\bf T}_{\rm R}:=n\frac{\partial \epsilon}{\partial{\bf \overline{e}}_i}
\otimes {\bf e}_i=n\frac{\partial \epsilon}{\partial{\bf F}},
\end{equation}
so that the Eshelby relation (\ref{eshelby_relation}) becomes %
\begin{equation}
{\bf c}_i := \frac{\partial( n \omega)}{\partial{\bf e}^i}=
n \omega\, {\bf e}_i-
{\bf T}_{\rm R}^\top{\bf F}{\bf e}_i-
({\bf g}^j\times{\bf e}_i)\times{\bf k}_j\label{eshelby_inc}
\end{equation}
due to the superposed deformation.

Thus  the expression (\ref{variation-0}) for the variation of the
energy functional becomes
\begin{align}
\delta\Omega =&
\int_{{\cal B}^\alpha\cup{\cal B}^\beta}
\big[
n(\frac{\partial \epsilon}{\partial\varrho}-\mu)\,
\delta\varrho +n(\frac{\partial \epsilon}{\partial s}-\theta_B)\,
\delta s  \notag\\
&\qquad\qquad\qquad
 +
({\bf c}_i
+\mbox{\Curl\,}{\bf k}_i)
\cdot\delta{\bf e}^i-(\Div\,{\bf T}_{\rm R})\cdot\delta{\bf f}
  \big]  \,dv
   \notag\\
&\qquad
-\int_{\cal S}\big( \ljump
 n\omega-{\bf n}_{_\mathcal{ S}}\cdot{\bf F}^\top{\bf T}_{\rm R}
{\bf n}_{_\mathcal{ S}}\rjump\, \delta r
+
{\bf n}_{_\mathcal{ S}}\!
\cdot\ljump{\delta}{\bf e}^i\times{\bf k}_i\rjump \notag\\
&\qquad \qquad\qquad
+\ljump{\bf T}_{\rm R}{\bf n}_{_\mathcal{
S}}\rjump\cdot\delta_\mathcal{S}{\bf f}
  \big) \,da =0,
\label{incoherent-inc}
\end{align}
where we have used the identity
\begin{equation}
-\ljump{\bf T}_{\rm R}{\bf n}_{_\mathcal{ S}}\cdot\delta{\bf f}\rjump=
-\ljump{\bf T}_{\rm R}{\bf n}_{_\mathcal{
S}}\rjump\cdot\delta_\mathcal{S}{\bf f}
+{\bf n}_{_\mathcal{ S}}\cdot\ljump{\bf F}^\top{\bf T}_{\rm R}
{\bf n}_{_\mathcal{ S}}\rjump\,\delta r,
\end{equation}
and  introduced
\begin{equation}
\delta_\mathcal{S}{\bf f}:=\delta{\bf f}+({\sf grad}_{{\bf n}_{\mathcal{ S}}}{\bf f})
\,\delta r =
\delta{\bf f}+{\bf F}{\bf n}_{_\mathcal{ S}}
\,\delta r,
\end{equation}
which is the
variation of the deformation
${\bf f}$ following the variation of the interface. Notice
that the
 continuity of ${\bf f}$ implies the continuity of
$\delta_\mathcal{S}{\bf f}$.  The difference between (\ref{incoherent-inc})
and (\ref{incoherent-0}) is due to the
superposed elastic deformation ${\bf f}$.

We take as the admissible class of independent variations
in bulk:
\begin{equation}
\delta\varrho,\quad
\delta s,\quad
\delta{\bf f},\quad
\delta{\bf e}^i.
\end{equation}
The introduction of the incremental deformation ${\bf f}$ provides
an additional admissible variation to (\ref{variations}).
This choice in (\ref{incoherent-inc})
leads to the following necessary conditions:\footnote{The derivatives
are, as before, with respect to the position {\bf x}, which is now in the
reference configuration for the incremental elastic deformation.}  \\
\begin{itemize}
\item[-]\emph{uniform chemical potential}
\begin{equation}
\frac{\displaystyle\partial \epsilon^\alpha}{\displaystyle\partial\varrho}
=\frac{\displaystyle\partial \epsilon^\beta}{\displaystyle\partial\varrho}
=\mu ; \label{inc_chemical_potential}\end{equation}
\item[-]\emph{uniform temperature}
\begin{equation}
\frac{\displaystyle \partial \epsilon^\alpha}{\displaystyle\partial s}=
\frac{\displaystyle \partial \epsilon^\beta}{\displaystyle\partial s}
=\theta_B ;\label{inc_temperature}\end{equation}
\item[-]\emph{Piola force balance for each phase}
\begin{equation}\Div{\bf T}_{\rm R}={\bf 0}; \label{inc_cauchy}
\end{equation}
\item[-]\emph{dislocation couple balance for each phase}
\begin{equation}
{\bf c}_i+{\Curl\,}{\bf k}_i ={\bf 0}.
\label{inc_couple}
\end{equation} \\[-5mm]
\end{itemize}
As in the previous unconstrained cases,
the couple balance (\ref{inc_couple}) implies immediately the
\emph{configurational force balance for each phase}
\begin{equation}
\Div\,{\bf c}_{i}=0,
\label{configurational_inc}
\end{equation}
but, importantly, it is \emph{not equivalent} to the force balance
(\ref{inc_cauchy}). The \emph{independent} conditions in the
bulk are  (\ref{inc_chemical_potential})--(\ref{inc_couple}).
In particular, there are two distinct mechanical equilibrium conditions:
the couple balance (\ref{inc_couple}) enforces mechanical equilibrium of the
dislocated intermediate region $\mathcal{B}$;  the force balance
(\ref{inc_cauchy}) enforces mechanical equilibrium of the elastically deformed
 region ${\bf f}(\mathcal{B})$.  Interestingly, both the intermediate and
the current configurations must be in mechanical equilibrium.\footnote{
Both (\ref{inc_cauchy}) and (\ref{inc_couple}) correspond to equations
derived variationally by Le \& Stumpf (1996) in a continuum theory
of dislocations with a postulated
elastic-plastic decomposition, though they were not
interpreted as the mechanical equilibrium conditions in the actual
and intermediate configurations.}

If we restrict the admissible lattice variations to the
holonomically constrained class
(i.e., the Born rule) then (\ref{configurational_inc}) replaces
(\ref{inc_couple}) as the mechanical equilibrium condition in the intermediate
configuration. Then both the configurational force balance 
(\ref{configurational_inc}) and the standard force balance (\ref{inc_cauchy})
are needed to enforce mechanical equilibrium.

At the  interface we take as the admissible independent variations
\begin{equation}
\delta{r}, \qquad
\delta_\mathcal{S}{\bf f},\qquad
({\delta}{\bf e}^i)^{\alpha}, \qquad
({\delta}{\bf e}^i)^{\beta}.
\end{equation}
They yield the following interface conditions:

\begin{itemize}
\item[-]\emph{continuity of  incremental normal Eshelby traction}
\begin{equation}
{\bf n}_{_\mathcal{ S}}\cdot\ljump n\omega{\bf 1}-{\bf F}^\top{\bf T}_{\rm R}
\rjump{\bf n}_{_\mathcal{ S}}=0;\label{norm_eshelby_inc}
\end{equation}
\item[-]\emph{continuity of incremental Piola traction}
\begin{equation}
\ljump {\bf T}_{\rm R}\rjump{\bf n}_{_\mathcal{ S}}={\bf 0};\label{piola_inc}
\end{equation}
\item[-]\emph{vanishing  of the individual tangential couples}
 %
%
\begin{equation}
{\sf P}({\bf k}_i)^{\alpha}={\bf 0}
\qquad \text{and} \qquad
{\sf P}
({\bf k}_i)^{\beta}={\bf 0}.
\label{inc_couple_traction} \end{equation} \\[-5mm]
\end{itemize}
Both (\ref {norm_eshelby_inc}) and (\ref{piola_inc}) are standard equilibrium
conditions for coherent interfaces, whereas (\ref{inc_couple_traction})
is  the condition for mechanical equilibrium for an incoherent interface.
Both sets of conditions arise because the interface is coherent under the
superposed elastic deformation, but incoherent in the intermediate
dislocated configuration. As in the bulk, we see that the interface must be in
equilibrium in both configurations when an elastic-plastic decomposition
is introduced.

If we restrict the lattice variations
in the intermediate configuration
to the holonomically constrained class,
then the \emph{vanishing of the individual
configurational tractions in the intermediate configuration} replaces the
equilibrium condition
(\ref{inc_couple_traction}). Note that  this configurational stress
differs from the incremental Eshelby stress, which arises from the
elastic part of the deformation.

\section{Discussion}

One way of introducing the notion of a discrete
 crystalline lattice into a macroscopic continuum model is  through 
additional fields that represent, for example,
the lattice basis vectors of the
crystalline structure at each point.
It is common however to eliminate these additional degrees of freedom
by means of the Born rule, which states that the
lattice vectors at each point deform with the
 gradient of the macroscopic deformation. Thus a crystalline body
 is reduced to an anisotropic elastic continuum, which remembers
 its lattice structure only through the symmetry of the constitutive
 relations.
 This approach fails when defects such as dislocations have to be accounted
for in the  model, since    the Born-rule no longer holds.
 In this case the macroscopic deformation gradient fails to describe 
  the true local ordering of the lattice.

In this work, we use the kinematic ideas of Davini and Parry on
defective crystals to construct a model of crystalline solid equilibrium. 
The basic assumption is that the defect density
 is sufficiently small, so that an average ordering is still recognizable locally.
 At each spatial point of the solid 
a lattice basis may therefore be assigned, but these
 local descriptions need not match globally to yield a smooth
 deformation of a perfect crystal. Dislocations are then viewed
as the obstruction to patching together such local lattice bases to a global
lattice basis.

   Using a  variational framework, we  obtain equilibrium
  equations for the bulk and for incoherent interfaces 
 in the presence of vacancies and
  dislocations without assuming the validity of the Born rule.  
   Although 
   the mechanical equilibrium equations involve 
   dislocation couples, indeed they also imply the more
   familiar equilibrium conditions ensuing from the Cauchy and 
   configurational force balances. We
  discuss how these equations simplify when the Born rule is imposed
  as a constraint on the possible lattice variations so that
 the defect density
  vanishes in bulk, and also when
   coherent boundaries are  considered. In these constrained cases, our results
  correspond to those previously derived by  Larch\'e \& Cahn (1978),
Leo \& Sekerka (1989),  and
 Cermelli \& Gurtin (1994). Importantly, our unconstrained results 
  provide a generalization to  
  their work, which only dealt with conventional elastic
 bodies, or, in our language,  holonomic microstructures.

  In order to compare to  common models involving
 an elastic-plastic decomposition of the strain, we also discuss an
  extended model, in which we superpose conventional   elastic
 deformations onto equilibrium states of a dislocated crystal.
  The resulting system for the bulk is composed of a
   force balance that enforces mechanical equilibrium of the
  current configuration, together with  a couple balance enforcing the
  mechanical equilibrium  of the dislocated intermediate
  configuration. Such bulk conditions have also been derived by
  Le \& Stumpf (1996) in a slightly different context.
  The  interfacial conditions are composed of
the standard  coherency relations 
   requiring the continuity of the Piola traction and the
  continuity of the usual Eshelby-stress traction, but  complemented
  by an incoherent relation requiring the vanishing of the individual
 tangential couples, which enforces the equilibrium of the
  interface with respect to the lattices of each phase in the intermediate 
  dislocated configuration.


\section*{Acknowledgements} This work was supported by the
MURST of Italy (Progetto `Metodi Matematici per la Scienza dei 
Materiali') and the EPSRC of the UK.


\section*{References}\frenchspacing

\begin{itemize}

\item[]{\hspace{-.75cm}}
Alexander, J.I.D.  \&  Johnson, W.C.   (1985) Thermomechanical
equilibrium in solid-fluid systems with curved interfaces.
\emph{J. Appl. Phys.} {\bf 58}, 816--824.

\item[]{\hspace{-.75cm}}
Besseling, J.F.  \& van der Giessen, E. (1994)
\emph{Mathematical Modelling of
Inelastic Deformation}, ch. 7. Chapman Hall, London.

\item[]{\hspace{-.75cm}}
Bilby, B.A. (1955) Types of dislocation source. In: \emph{Report Conf.
Defects in Crystalline Solids} (Bristol, 1954), The Physical Society, London,
124--133.

\item[]{\hspace{-.75cm}}
Bilby, B.A., Gardner, L.R.T. \& Stroh, A.N. (1957) Continuous distributions
of dislocations and the theory of plasticity. In: \emph{Extrait des Actes  du
IX$^e$ Congr\`es International de M\'ecanique Appliqu\'ee}, Brussels, 35--44.

\item[]{\hspace{-.75cm}}
Bilby, B.A. \& Smith, E. (1956) Continuous distributions of dislocations III.
\emph{Proc. R. Soc. Lond.} {\bf A236}, 481--505.

\item[]{\hspace{-.75cm}}
Cahn,  J.W. (1980) Surface stress and the chemical equilibrium of small
crystals---I.\ The case of the isotropic surface.
{\it Acta Metall.} {\bf 28}, 1333--1338.

\item[]{\hspace{-.75cm}}
Cahn, J.W.  \&  Larch\'e, F.  (1982)
Surface stress and the chemical equilibrium of small crystals---II.\
Solid particles embedded in a solid matrix.
{\it Acta Metall.} {\bf 30}, 51--56.


\item[]{\hspace{-.75cm}}
Cermelli, P. \& Gurtin, M.E. (1994)   The dynamics
of solid-solid phase transitions. 2. Incoherent interfaces. \emph{Arch.
Rational Mech. Anal.} {\bf 127},  41--99.


\item[]{\hspace{-.75cm}}
Cermelli, P. \& Sellers, S.  (1998)
 On the nonlinear mechanics of bravais crystals
with continous distributions of defects. \emph{Mathematics and
Mechanics of Solids} {\bf 3}, 331--358.

\item[]{\hspace{-.75cm}}
Davini, C. (1986)  A proposal for a continuum theory of defective
crystals. \emph{Arch. Rational Mech. Anal.} {\bf 96},  295--317.

\item[]{\hspace{-.75cm}}
Davini, C. \& Parry, G.P. (1989)
On defect preserving deformations in crystals.
{\it Int. J. Plasticity} {\bf 5},  337--369.

\item[]{\hspace{-.75cm}}
Davini, C. \& Parry, G.P. (1991)  A complete list of
invariants for defective crystals.  \emph{Proc.  R. Soc.  Lond. }
{\bf A 432}, 341--365.

\item[]{\hspace{-.75cm}}
D{\l}u\.zewski, P.  (1996) On the geometry and continuum
thermodynamics of movement of structural defects.
\emph{Mechanics of Materials}, \textbf{ 22}, 23--41.


\item[]{\hspace{-.75cm}}
 Ericksen, J.L. (1984) The Cauchy and Born hypothesis
for crystals. In: Gurtin, M.E. (ed.),
\emph{Phase Transformations and Material
Instabilities in Solids},
Academic Press, New York, 61--77.

\item[]{\hspace{-.75cm}}
Ericksen, J.L. (1997)
Equilibrium theory for X-ray observations of crystals.
\emph{Arch. Rational Mech. Anal.}  {\bf 139}, 181--200.

\item[]{\hspace{-.75cm}}
Ericksen, J.L. \& Rivlin, R.S. (1954) Large elastic deformations of
homogeneous anisotropic
materials. \emph{J. Rational Mech. Anal.} {\bf 3}, 281--301.

\item[]{\hspace{-.75cm}}
Ericksen, J.L. \& Truesdell, C. (1958) Exact theory of stress and strain
in rods and shells. \emph{Arch. Rational Mech. Anal.}
{\bf 1}, 295--323.

\item[]{\hspace{-.75cm}}
Eshelby, J.D.  (1951) The force on an elastic singularity.
\emph{Phil. Trans. R. Soc. Lond.} {\bf A 244},  87--112.

\item[]{\hspace{-.75cm}}
Eshelby, J.D. (1970) Energy relations and the energy momentum tensor in
continuum mechanics. In: Kanninen, M.,
Adler, W., Rosenfield, A., \& Jaffee, R. (eds.),
\emph{Inelastic Behavior of Solids},  McGraw-Hill, New York,
 77--115.

\item[]{\hspace{-.75cm}}
Fonseca, I. \& Parry, G. (1992)
Equilibrium configurations of defective crystals.
\emph{Arch. Rational Mech. Anal.} {\bf 120}, 
245--283. 

\item[]{\hspace{-.75cm}}
Fox, N. (1966) A continuum theory of dislocations for single crystals.
\emph{J. Inst. Maths. Applics.} {\bf 2}, 285--298.

\item[]{\hspace{-.75cm}}
Fox, N.  (1970) A dislocation theory for oriented continua.
In: Kanninen, M.,
Adler, W., Rosenfield, A., \& Jaffee, R. (eds.),
\emph{Inelastic Behavior of Solids}, McGraw-Hill, New York,
349--361.

\item[]{\hspace{-.75cm}}
Gibbs, J.W. (1878) On the equilibrium of heterogeneous substances,
\emph{Trans. Connecticut Acad.} {\bf 3} 108--248; reprinted in
Gibbs, J.W (1961)
\emph{The Scientific Papers of J.W. Gibbs}, vol. 1, Dover, New York.

\item[]{\hspace{-.75cm}}
Grinfel'd, M.A. (1981) On heterogeneous equilibrium of nonlinear
elastic phases and chemical potential tensors. \emph{Lett. Appl. Engng.
Sci.} {\bf 19}, 1031--1039.

\item[]{\hspace{-.75cm}}
Gurtin, M.E.  (1993) The dynamics of solid-solid phase transitions.
1. Coherent interfaces. \emph{Arch. Rational Mech. Anal.} {\bf 123},
 305--335.

\item[]{\hspace{-.75cm}}
Gurtin, M.E. (1995) The nature of configurational forces.
\emph{Arch. Rational Mech. Anal}. {\bf 131},  67--100.

\item[]{\hspace{-.75cm}}
Johnson, W.C.  \& Alexander, J.I.D.  (1986)
 Interfacial conditions for thermomechanical
equilibrium in two-phase crystals. \emph{J. Appl. Phys.} {\bf 59},
2735--2746.

\item[]{\hspace{-.75cm}}
Kaganova, I.M.  \& Roitburd, A.L.  (1988) Equilibrium between
elastically-interacting phases
[English translation]. \emph{Sov. Phys. JETP}
{\bf 67}, 1173--1183.

\item[]{\hspace{-.75cm}}
Kosevich, A.M.  (1962) The deformation field in an isotropic
elastic medium containing moving dislocations [English
Translation]. \emph{Sov. Phys. JETP}  {\bf 15}, 108--115.

\item[]{\hspace{-.75cm}}
Kr\"oner, E. (1960) All\-ge\-meine Kon\-tinuums\-theorie der
Ver\-setz\-ung\-en und
Ei\-gen\-spann\-ung\-en. \emph{Arch. Rational Mech. Anal.} {\bf 4},
 273--334.

\item[]{\hspace{-.75cm}}
Larch\'e, F.C.  \& Cahn, J.W. (1978)
Thermomechanical equilibrium of multiphase
solids under stress. \emph{Acta Metall.} {\bf 26},  1579--1589.

\item[]{\hspace{-.75cm}}
Larch\'e, F.C. \& Cahn,  J.W.  (1985) The interaction of composition
and stress in crystalline solids. \emph{Acta Metall.} {\bf 33}, 331--357.

\item[]{\hspace{-.75cm}}
Le, K.C. \& Stumpf, H.  (1996) Nonlinear continuum theory of dislocations.
\emph{Int. J. Engng. Sci.} {\bf 34}, 339--358.

\item[]{\hspace{-.75cm}}
Lee, E.H.  (1969) Elastic-plastic deformation at finite strains.
\emph{J. App. Mechanics} {\bf 36}, 1--6.

\item[]{\hspace{-.75cm}}
Leo, P.H.  \&  Sekerka, R.F. (1989)
The effect of surface stress on crystal-melt and crystal-crystal
equilibrium. \emph{Acta Metall.} {\bf 37},  3119--3138.

\item[]{\hspace{-.75cm}}
Leo, P.H.  \& Hu, J.  (1995) A continuum description of partially coherent
interfaces. \emph{Continuum Mech. Thermodyn.} {\bf 7}, 39--56.

\item[]{\hspace{-.75cm}}
Maugin, G.  (1993) \emph{Material Inhomogeneities in Elasticity},
Chapman Hall, London.

\item[]{\hspace{-.75cm}}
Mullins, W.W. (1984) Thermodynamic equilibrium of a crystalline sphere
in a fluid. {\it J. Chem.  Phys.} {\bf 81}, 1436--1442.

\item[]{\hspace{-.75cm}}
Mullins, W.W.  \& Sekerka, R.F. (1985)
On the thermodynamics of crystalline solids. \emph{J. Chem. Phys.}
{\bf 82}, 5192--5202.


\item[]{\hspace{-.75cm}}
Naghdi, P.M.  \& Srinivasa, A.R. (1993a) A dynamical theory
of structured solids: I. Basic development.
\emph{Phil. Trans. R. Soc. Lond.}  {\bf  A 345}, 425--458.

\item[]{\hspace{-.75cm}}
Naghdi, P.M. \& Srinivasa,  A.R. (1993b) A dynamical theory
of structured solids: II. Special constitutive equations and special
cases of the theory.
\emph{Phil. Trans. R. Soc. Lond. } {\bf A 345}, 459--476.

\item[]{\hspace{-.75cm}}
Naghdi, P.M.   \& Srinivasa, A.R. (1994a) Characterization of
dislocations and their influence on plastic deformation in single crystals.
\emph{Int. J. Engng. Sci.}
{\bf 32}, 1157--1182.

\item[]{\hspace{-.75cm}}
Naghdi, P.M.   \& Srinivasa, A.R. (1994b)
Some general results in the theory of
crystallographic slip. \emph{Z. angew. Math. Phys.} {\bf 45}, 687--732.

\item[]{\hspace{-.75cm}}
Nye, J.F. (1953) Some geometrical relations in dislocated solids.
\emph{Acta Metall.} {\bf 1}, 153--162.



\item[]{\hspace{-.75cm}}
Robin, P.-Y.  (1974) Thermodynamic equilibrium across a coherent interface
in a stressed crystal. \emph{Am. Mineral.} {\bf 59}, 1286--1298.

\item[]{\hspace{-.75cm}}
Toupin, R.A. (1968) Dislocated and oriented media. In: Kr{\"o}ner, E. (ed.)
 \emph{Mechanics of
Generalized Continua},
 Springer, Berlin/Heidelberg, 126--140.

\item[]{\hspace{-.75cm}}
Truesdell, C.  \& Toupin, R.  (1960) The classical field theories.
In: Fl{\"u}gge, S. (ed.)
\emph{Handbuch der Physik}, vol. III/1, Springer, Berlin.

\item[]{\hspace{-.75cm}}
Zanzotto, G. (1992) On the material symmetry group of elastic crystals
and the Born rule. \emph{Arch. Rational Mech. Anal.} {\bf 121}, 1--36.

\end{itemize}


\end{document}